\documentclass{llncs}
\usepackage{lmodern}
\usepackage{amsmath,amssymb,color,bussproofs,epsfig,graphicx,verbatim,bm,galois,xcolor,xspace,todonotes,turnstile}
\usepackage{microtype}
\usepackage[pdftex,colorlinks=true,bookmarks=false,linkcolor=blue,citecolor=blue,urlcolor=blue] {hyperref}
\usepackage{colorbar}
\usepackage{multirow}
\usepackage{hhline}
\usepackage{etex}
\usepackage[all]{xy}
\usepackage{galois}
\usepackage{subcaption}
\usepackage{galois}
\usepackage{tikz}
\usetikzlibrary{arrows,automata}

\newcommand{\AP}{{\textit{AP}}\xspace}

\newcommand{\trans}{\textit{trans}\xspace}
\newcommand{\FeatExp}{\textit{FeatExp}}
\newcommand{\true}{\textit{true}}
\newcommand{\false}{\textit{false}}
\newcommand{\ttv}{\textit{tt}}
\newcommand{\ffv}{\textit{ff}}

\newcommand{\U}{\textsf{U}}
\newcommand{\V}{\textsf{V}}
\newcommand{\may}{\textrm{may}}
\newcommand{\must}{\textrm{must}}

\newcommand{\vending}{\textsc{VendMach}}

\newcommand{\sbr}[1]{\lbrack \! \lbrack #1 \rbrack \! \rbrack}

\newcommand{\poset}[2]{\ensuremath{\langle{#1},{#2}\rangle}}

\newcommand{\smv}{\textsc{NuSMV}}
\newcommand{\fsmv}{f\textsc{NuSMV}}

\newcommand{\elevator}{\textsc{Elevator}}

\newcommand{\Ff}{\mathbb{F}}
\newcommand{\Fff}{\mathcal{F}}
\newcommand{\Mmm}{\mathcal{M}}

\newcommand{\Ttt}{\mathcal{T}}
\newcommand{\Kk}{\mathbb{K}}

\newcommand{\joinasym}{\ensuremath{\bm{\alpha}^{\textnormal{\textrm{join}}}}}
\newcommand{\joingsym}{\ensuremath{\bm{\gamma}^{\textnormal{\textrm{join}}}}}

\definecolor{redgray}{rgb}{0.8,0.2,0.2}
\definecolor{greengray}{rgb}{0.2,0.6,0.2}
\definecolor{lightgray}{rgb}{0.8,0.8,0.8}
\definecolor{darkgreen}{rgb}{0.0,0.5,0.0}

\newcommand{\sr}{\textit{r}}

\newcommand{\ff}{\textcolor{blue}{\ensuremath{f}}}
\newcommand{\fc}{\textcolor{brown}{\ensuremath{c}}}

\newcommand{\TR}[1]{%
  \raisebox{-.3mm}{\ensuremath{\xrightarrow{\smash{\!#1}}}}}

\title{Variability Abstraction and Refinement for Game-based Lifted Model Checking of full CTL (Extended Version)
}

\author{  Aleksandar S. Dimovski\inst{1} \and Axel Legay\inst{2} \and Andrzej Wasowski\inst{3}
}

\institute{  Mother Teresa University, 12 Udarna Brigada 2a, 1000 Skopje, Mkd \and
            UCLouvain, Belgium and IRISA/NRIA Rennes, France    \and
     IT University of Copenhagen, Rued Langgaards Vej 7, 2300 Copenhagen, Denmark
}

\begin{document}
\maketitle

\begin{abstract}

Variability models allow effective building
of many custom model variants for various configurations. Lifted
model checking for a variability model is capable of verifying all its variants
simultaneously in a single run by exploiting the similarities between the
variants.
The computational cost of lifted
model checking still greatly depends on the number of variants (the size of configuration space), which is often huge.

One of the most promising approaches to fighting the configuration space explosion problem
in lifted model checking
are \emph{variability abstractions}.
In this work, we define a novel game-based approach for variability-specific abstraction and refinement for
 lifted model checking of the full CTL, interpreted over 3-valued semantics.
We propose a direct algorithm for solving a 3-valued (abstract) lifted model checking game.
In case the result of model checking an abstract variability model is indefinite, we suggest
a new notion of refinement, which eliminates indefinite results.
This provides an iterative incremental variability-specific abstraction and refinement
framework, where refinement is applied only where indefinite results exist and definite
results from previous iterations are reused. 

\end{abstract}

\section{Introduction}\label{sec:introduction}

\noindent
Software Product Line (SPL) \cite{SPL} is an efficient method for systematic development
of a family of related models, known as \emph{variants} (\emph{valid products}), from a common code base.
Each variant is specified in terms of \emph{features} (static configuration options) selected for that particular
variant.
SPLs are particulary popular in the embedded and critical system domains (e.g. cars, phones, avionics, healthcare).

Lifted 
model checking \cite{model-checking-spls,classen-model-checking-spls-icse11} is a useful approach for verifying
properties of variability models (SPLs).
Given a variability model and a specification, 
the lifted model checking algorithm, unlike the standard non-lifted one, returns precise conclusive results for all individual variants,
that is, for each variant it reports whether it satisfies or violates the specification. 
The main disadvantage of lifted model checking is the \emph{configuration space explosion problem}, which
refers to the high number of variants in the variability model. 
In fact, exponentially many variants can be derived from only few configuration options (features).
One of the most successful approaches to fighting the configuration space explosion are
so-called \emph{variability abstractions} \cite{spin15,sttt16,DBLP:conf/birthday/DimovskiW17,fase18}. They hide some of the configuration details,
so that many of the concrete configurations become indistinguishable and can be collapsed into a
single abstract configuration (variant).
This results
in smaller abstract variability models with a smaller number of abstract configurations, such that
each abstract configuration corresponds to some subset of concrete configurations.
In order to be conservative w.r.t. the full CTL temporal logic, abstract variability models have two types of transitions:
\emph{may-transitions} which represent possible transitions in the concrete model, and
\emph{must-transitions} which represent the definite transitions in the concrete model.
May and must transitions  correspond to over and under approximations, and are needed in order
to preserve universal and existential CTL properties, respectively.

Here we consider the 3-valued semantics for interpreting CTL formulae over abstract variability models.
This semantics evaluates a formula on an abstract model to either \emph{true}, \emph{false}, or \emph{indefinite}.
Abstract variability models are designed to be conservative for both \emph{true} and \emph{false}.
However, the \emph{indefinite} answer gives no information on the value of the formula on the concrete model.
In this case, a refinement is needed in order to make the abstract models more precise.

The technique proposed here significantly extends the scope of existing automatic variability-specific abstraction refinement procedures \cite{DBLP:conf/sigsoft/CordyHLSDL14,fase17},
which currently support the verification of universal LTL properties only.
They use conservative variability abstractions to construct over-approximated abstract variability
models, which preserve LTL properties.
If a spurious counterexample (introduced due to the abstraction) is found in the abstract model,
the procedures \cite{DBLP:conf/sigsoft/CordyHLSDL14,fase17} use
Craig interpolation to extract relevant information from it in
order to define the refinement of abstract models.
Variability abstractions
that preserve all (universal and existential) CTL properties have been previously introduced \cite{fase18}, but
without an automatic mechanism for constructing them and no notion of
refinement. The abstractions \cite{fase18} has to be constructed manually
before verification.
In order to make the entire verification procedure automatic, we need to develop
an abstraction and refinement framework for CTL properties.

In this work, we propose the first variability-specific abstraction refinement procedure
for automatically verifying arbitrary formulae of CTL. To achieve this aim,
model checking \emph{games} \cite{DBLP:series/txcs/Stirling01,DBLP:journals/tocl/ShohamG07,DBLP:journals/iandc/ShohamG10} represent
the most suitable framework for defining the refinement.
In this way, we establish a brand new connection between games and family-based (SPL) model checking. 
The refinement is defined by finding the reason for the indefinite result of an algorithm
that solves the corresponding model checking game, which is played by two players: Player$\, \forall$ and Player$\, \exists$.
The goal of Player$\, \forall$ is either to refute the formula $\Phi$ on an abstract model $\Mmm$ or to prevent Player$\, \exists$ from verifying it.
Similarly, the goal of Player$\, \exists$ is either to verify $\Phi$ on $\Mmm$ or to prevent Player$\, \forall$ from refuting it.
The game is played on a \emph{game board}, which consists of configurations of the form $(s,\Phi')$
where $s$ is a state of the abstract model $\Mmm$ and $\Phi'$ is a subformula of $\Phi$, such that the value of
$\Phi'$ in $s$ is relevant for determining the final model checking result.
The players make moves between configurations in which they try to verify or refute $\Phi'$ in $s$.
All possible plays of a game are captured in the game-graph, whose nodes are the elements of the game board and
whose edges are the possible moves of the players.
The model checking game is solved via a coloring algorithm which colors each node $(s,\Phi')$
in the game-graph by $T$, $F$, or $?$ iff the value of $\Phi'$ in $s$ is \true, \false, or indefinite, respectively.
Player$\, \forall$
has a winning strategy at the node $(s,\Phi')$ iff the node is colored by $F$ iff $\Phi'$ does not hold in $s$,
and Player$\, \exists$ has a winning strategy at $(s,\Phi')$ iff the node is colored by $T$ iff $\Phi'$ holds in $s$.
In addition, it is also possible that neither of players has a winning strategy, in which case the node is colored by $?$ and the value
of $\Phi'$ in $s$ is indefinite.

When the game results in a \emph{tie} with an indefinite answer, we want to refine the abstract model.
We can find the reason for the tie by examining the part of the game-graph which has indefinite results.
We choose then a refinement criterion, which splits abstract configurations so that the new, refined
abstract configurations represent smaller subsets of concrete configurations.

\vspace{-1mm}
\section{Background} \label{sec:background}


\paragraph{\textbf{Variability Models.}}
Let $\Ff = \{A_1, \ldots, A_n\}$ be a finite set of Boolean variables representing
the features available in a variability model.
A specific subset of features, $k \subseteq \Ff$, known as \emph{configuration}, specifies
a \emph{variant} (valid product) of a variability model.
We assume that only a subset \(\Kk \subseteq 2^{\Ff}\) of configurations are \emph{valid}.
An alternative representation of configurations is based upon
propositional formulae. Each configuration $k \in \Kk$ can be represented by a formula:
$k(A_1) \land \ldots \land k(A_n)$,
where $k(A_i) = A_i$ if $A_i \in k$, and $k(A_i) = \neg A_i$ if $A_i \notin k$ for $1 \leq i \leq n$.
We will use both representations interchangeably.

We use \emph{transition systems} (TS) to describe behaviors of single-systems.
\begin{definition}
A transition system (TS) is a tuple $\Ttt=(S,Act,trans,I,AP,L)$, where $S$ is a set of states;
$Act$ is a set of actions; $trans \subseteq S \times Act \times S$ is a transition
relation which is \emph{total}, so that for each state there is an outgoing transition;
$I \subseteq S$ is a set of initial states; $AP$ is a set of atomic propositions;
and $L : S \to 2^{AP}$ is a labelling function specifying which propositions hold in a state.  We write $s_1 \TR {~\lambda~} s_2$ whenever \((s_1,\lambda,s_2) \in \trans\).
\end{definition}
An \emph{execution} (behaviour) of a TS $\Ttt$ is an \emph{infinite} sequence $\rho = s_0 \lambda_1 s_1 \lambda_2 \ldots$
with $s_0 \in I$ such that $s_i \stackrel{\lambda_{i+1}}{\longrightarrow} s_{i+1}$ for all $i \geq 0$.
The \emph{semantics} of the TS $\Ttt$, denoted as $\sbr{\Ttt}_{TS}$, is the set of its executions.

A \emph{featured transition system} (FTS) is a particular instance of a variability model,
which describes the behavior of a whole family of systems 
in a single monolithic description, where the transitions are guarded by a \emph{presence condition} that identifies the variants they belong to.  The presence conditions $\psi$ are drawn from the set of feature expressions, $\FeatExp(\Ff)$, which are  propositional logic formulae over $\Ff$:
$  \psi ::= \true \mid A \in \Ff \mid \neg \psi \mid \psi_1 \land \psi_2$.
We write $\sbr{\psi}$ to denote the set of configurations from $\Kk$ that satisfy $\psi$, i.e.\ $k \in \sbr{\psi}$ iff $k \models \psi$.


\begin{definition}
A featured transition system (FTS) represents a tuple
$\Fff=(S,Act,trans,I,AP,L,\Ff,\Kk,\delta)$, where $S, Act, trans, I, AP$, and $L$
form a TS; $\Ff$ is the set of available features; $\Kk$ is a set of valid configurations; and
$\delta: trans \!\to\! \FeatExp(\Ff)$ is a total function decorating transitions with presence conditions. 
\end{definition}
The \emph{projection} of an FTS $\Fff$ to a configuration $k \in \Kk$, denoted as $\pi_k(\Fff)$, is the TS
$(S,Act,trans',I,AP,L)$, where $trans'=\{ t \in trans \mid k \models \delta(t) \}$.
We lift the definition of \emph{projection} to sets of configurations \(\Kk' \!\subseteq\! \Kk\),
 denoted as $\pi_{\Kk'}(\Fff)$, by keeping the transitions admitted by at least one of the configurations in $\Kk'$.
That is, $\pi_{\Kk'}(\Fff)$, is the FTS
$(S,Act,trans',I,AP,L,\Ff,\Kk',\delta')$, where $trans'=\{ t \in trans \mid \exists k \in \Kk'. k \models \delta(t) \}$
and $\delta' = \left.\delta \right|_{trans'}$ is the restriction of $\delta$ to $trans'$.
The \emph{semantics} of an FTS $\Fff$, denoted as $\sbr{\Fff}_{FTS}$, is the union of behaviours
of the projections on all valid variants $k \in \Kk$, i.e.\ $\sbr{\Fff}_{FTS} = \cup_{k \in \Kk} \sbr{\pi_k(\Fff)}_{TS}$.

\emph{Modal transition systems} (MTSs) \cite{DBLP:conf/lics/LarsenT88} are a generalization of transition systems
equipped with two transition relations: \emph{must} and \emph{may}. The former (must) is used to specify the required behavior, while the latter (may) to specify the allowed behavior of a system.
We will use MTSs for representing abstractions of FTSs.
\begin{definition}
A modal transition system (MTS) is represented by a tuple $\Mmm=(S,Act,trans^{\may},trans^{\must},I,AP,L)$, where
$trans^{\may} \subseteq S \times Act \times S$ describe may transitions of $\Mmm$;
$trans^{\must} \subseteq S \times Act \times S$ describe must transitions of $\Mmm$,
such that $trans^{\may}$ is total and $trans^{\must} \subseteq trans^{\may}$.
\end{definition}
A \emph{may-execution} in $\Mmm$ is an execution (infinite sequence) with all its transitions in $trans^{\may}$; whereas
a \emph{must-execution} in $\Mmm$ is a maximal sequence with all its transitions in $trans^{\must}$, which cannot be
extended with any other transition from $trans^{\must}$. Note that since $trans^{\must}$ is not
necessarily total, must-executions can be finite.
We use $\sbr{\Mmm}_{MTS}^{\may}$ (resp., $\sbr{\Mmm}_{MTS}^{\must}$) to denote the set of all may-executions
(resp., must-executions) in $\Mmm$ starting in an initial state.

\begin{figure}[t]
\centering
\begin{minipage}[b]{.32\textwidth}
\centering
\begin{tikzpicture}[->,>=stealth',shorten >=0.9pt,auto,node distance=1.4cm, semithick]
  \tikzstyle{every state}=[minimum size=.2pt,initial text={{}}]

  \node[initial below,state] (A)     [label=above:{\tiny $\neg \textit{r}$}]  {\tiny $s_0$};
  \node[state]         (B) [right of=A, label=above:{\tiny $\neg \textit{r}$}] {\tiny $s_1$};
  \node[state]         (C) [right of=B, label=above:{\tiny $\textit{r}$}] {\tiny $s_2$};

  \path[font=\tiny] (A) edge node {$\textit{pay} /\!\neg \ff$} (B)
        (B) edge node {$\textit{drink}$} (C)
        (C) edge [bend left] node {$\textit{take}$} (A);

  \path[draw=blue,fill=blue,font=\tiny]      (A) edge [bend left=75]
  node[above,sloped] {{\color{blue}$\textit{free} /\! \ff$}} (C);

  \path[draw=brown,fill=brown,font=\tiny]
        (B) edge [bend right=45] node[above,sloped] {{\color{brown}$\textit{cancel} /\! c$}} (A);

\end{tikzpicture}
\vspace{-2.5mm}
\caption{\small \vending}
\label{fig:FTS}
\end{minipage}%
\begin{minipage}[b]{.33\textwidth}
\centering
\begin{tikzpicture}[->,>=stealth',shorten >=0.9pt,auto,node distance=1.4cm,
                    semithick]
  \tikzstyle{every state}=[minimum size=.15pt,initial text={{}}]

  \node[initial below,state] (A)     [label=above:{\tiny $\neg \textit{r}$}]  {\tiny $s_0$};
  \node[state]         (B) [right of=A, label=above:{\tiny $\neg \textit{r}$}] {\tiny $s_1$};
  \node[state]         (C) [right of=B, label=above:{\tiny $\textit{r}$}] {\tiny $s_2$};

  \path[font=\tiny] (A) edge node {$\textit{pay}$} (B)
        (B) edge node {$\textit{drink}$} (C)
        (C) edge [bend left] node {$\textit{take}$} (A);

\end{tikzpicture}
\vspace{-2.5mm}
\caption{$\pi_{\emptyset}(\small \vending)$}
\label{fig:variant1}
\end{minipage}
\begin{minipage}[b]{.34\textwidth}
\centering
\begin{tikzpicture}[->,>=stealth',shorten >=0.9pt,auto,node distance=1.4cm, semithick]
  \tikzstyle{every state}=[minimum size=.2pt,initial text={{}}]

  \node[initial below,state] (A)     [label=above:{\tiny $\neg \textit{r}$}]  {\tiny $s_0$};
  \node[state]         (B) [right of=A, label=above:{\tiny $\neg \textit{r}$}] {\tiny $s_1$};
  \node[state]         (C) [right of=B, label=above:{\tiny $\textit{r}$}] {\tiny $s_2$};

  \path[font=\tiny]
        (B) edge node {$\textit{drink}$} (C)
        (C) edge [bend left] node {$\textit{take}$} (A);

  \path[draw, dashed, font=\tiny]
  (A) edge node {$\textit{pay}$} (B)
  (A) edge [bend left=75]  node[above,sloped] {{$\textit{free}$}} (C)
  (B) edge [bend right=45] node[above,sloped] {{$\textit{cancel}$}} (A);

\end{tikzpicture}
\vspace{-2.5mm}
\caption{\small \joinasym(\vending)}
\label{fig:abstractFTS}
\end{minipage}%
\end{figure}

\begin{example} \label{exp:1}
Throughout this paper, we will use a beverage vending machine as a running example \cite{model-checking-spls}.
Figure~\ref{fig:FTS} shows the FTS of a \vending \, family.
It has two features, and each of them is assigned an identifying letter and a color. The features are:
\texttt{CancelPurchase} (\fc, \textcolor{brown}{in brown}), for canceling a purchase after a coin is entered; and
\texttt{FreeDrinks} (\ff, \textcolor{blue}{in blue}) for offering free drinks.
Each transition is labeled by an \emph{action} followed by a \emph{feature expression}. 
For instance, the transition $s_0$ \textcolor{blue}{\TR{\textit{free}/\!f}} $s_2$ is included in variants where the feature \textcolor{blue}{$f$} is enabled.
For clarity, we omit to write the presence condition \true\, in transitions.
There is only one atomic proposition $\texttt{served} \in AP$, which is abbreviated as $\textit{r}$.
Note that $\textit{r} \in L(s_2)$, whereas $\textit{r} \not\in L(s_0)$ and $\textit{r} \not\in L(s_1)$.

By combining various features, a number of variants of this \vending\,
can be obtained.
The set of valid configurations is:
  \noindent$\displaystyle\Kk^{\textsc{VM}} \!=\! \{ \emptyset,\linebreak[0] \{ \fc\},\linebreak[0] \{ \ff\},\linebreak[0] \{ \fc,\linebreak[0] \ff \} \}$
  (or, equivalently $\displaystyle\Kk^{\textsc{VM}} \!=\! \{ \neg \fc \land\! \neg \ff,\linebreak[0] \fc \land\! \neg \ff,\linebreak[0] \neg \fc \land\! \ff,\linebreak[0] \fc \land\! \ff \}$).
Figure~\ref{fig:variant1} shows a basic version of \vending\, that only serves a drink,
described by the configuration: $\emptyset$ (or, as formula $\neg \fc \land\! \neg \ff$). 
It takes a coin, serves a drink, opens a compartment
so the customer can take the drink.
Figure~\ref{fig:abstractFTS} shows an MTS, where must transitions are denoted by solid lines, while
may transitions by dashed lines.
\qed
\end{example}

\paragraph{\textbf{CTL Properties.}}
We present Computation Tree Logic (CTL)\,\cite{katoen-beier} for specifying system properties.
CTL state formulae $\Phi$ are given by:
\[
\Phi ::= \true \mid \false \mid l \mid \Phi_1 \land \Phi_2 \mid \Phi_1 \lor \Phi_2 \mid  A \phi \mid E \phi, \qquad
\phi ::=  \bigcirc \Phi \mid \Phi_1 \U \Phi_2  \mid \Phi_1 \V \Phi_2
\]
where $l \in Lit = \AP \cup \{\neg a \mid a \in \AP\}$ and $\phi$ represent CTL path formulae.
Note that the CTL state formulae $\Phi$ are given in negation normal form
($\neg$ is applied only to atomic propositions).
The path formula $\bigcirc \Phi$ can be read as ``in the next state $\Phi$'',
$\Phi_1 \U \Phi_2$ can be read as ``$\Phi_1$ until $\Phi_2$'', and its dual
$\Phi_1 \V \Phi_2$ can be read as ``$\Phi_2$ while not $\Phi_1$'' (where $\Phi_1$ may never hold).

 We assume the standard CTL semantics over TSs is given \cite{katoen-beier} (see also Appendix~\ref{app:ctl}). 
  We write $[\Ttt \models \Phi]=\ttv$ to denote that $\Ttt$ satisfies the formula $\Phi$, whereas
  $[\Ttt \models \Phi]=\ffv$ to denote that $\Ttt$ does not satisfy $\Phi$.

We say that an FTS $\Fff$ satisfies a CTL formula $\Phi$, written $[\Fff \models \Phi]=\ttv$, iff all its valid variants satisfy the formula, i.e.\
$  \forall k\!\in\!\Kk. \, [\pi_k(\Fff) \models \Phi]=\ttv$. Otherwise, we say
$\Fff$ does not satisfy $\Phi$, written $[\Fff \models \Phi]=\ffv$.
In this case, we also want to determine a non-empty set of violating variants $\Kk' \subseteq \Kk$, such that
$\forall k'\!\in\!\Kk'. \,[\pi_{k'}(\Fff) \models \Phi]=\ffv$ and $\forall k\!\in\!\Kk \backslash \Kk'. \,[\pi_k(\Fff) \models \Phi]=\ttv$.

We define the 3-valued semantics of CTL over an MTSs
 $\Mmm$ slightly differently from the semantics for TSs. 
 A CTL state formula $\Phi$ is satisfied in a state $s$ of an MTS $\Mmm$, denoted $[\Mmm,s \models^3 \Phi]$,
iff ($\Mmm$ is omitted when clear from context): \footnote{See Appendix~\ref{app:ctl} for definitions of $[s \models^3 \Phi_1 \lor \Phi_2]$, $[\rho \models^3 \bigcirc \Phi]$, and $[\rho \models^3 \!(\Phi_1 \V \Phi_2)]$.}
\begin{description}
\item[(1)] $[s \models^3 a] = \left.\begin{cases} \ttv, & \text{if } a \in L(s) \\ \ffv, & \text{if } a \not\in L(s) \end{cases}\right.$, \quad $[s \models^3 \neg a] = \left.\begin{cases} \ttv, & \text{if } a \not\in L(s) \\ \ffv, & \text{if } a \in L(s) \end{cases}\right.$
\item[(2)] $[s \models^3 \Phi_1 \land \Phi_2] = \left.\begin{cases} \ttv, & \text{if } [s \models^3 \Phi_1]=\ttv \text{ and } [s \models^3 \Phi_2]=\ttv \\ \ffv, & \text{if } [s \models^3 \Phi_1]=\ffv \text{ or } [s \models^3 \Phi_2]=\ffv \\ \bot, & \text{otherwise} \end{cases}\right.$ \\
\item[(3)] $[s \models^3 A \phi] = \left.\begin{cases} \ttv, & \text{if } \forall \rho \in \sbr{\Mmm}^{\may,s}_{MTS}. \, [\rho \models^3 \phi] = \ttv \\ \ffv, & \text{if } \exists \rho \in \sbr{\Mmm}^{\must,s}_{MTS}. \, [\rho \models^3 \phi] = \ffv \\ \bot, & \text{otherwise} \end{cases}\right.$ \\
$[s \models^3 E \phi] = \left.\begin{cases} tt, & \text{if } \exists \rho \in \sbr{\Mmm}^{\must,s}_{MTS}. \, [\rho \models^3 \phi] = \ttv \\ \ffv, & \text{if } \forall \rho \in \sbr{\Mmm}^{\may,s}_{MTS}. \, [\rho \models^3 \phi] = \ffv \\ \bot, & \text{otherwise} \end{cases}\right.$
\end{description}
where $\sbr{\Mmm}_{MTS}^{\may,s}$ (resp., $\sbr{\Mmm}_{MTS}^{\must,s}$) denotes the set of all may-executions
(must-executions) starting in the state $s$ of $\Mmm$.
Satisfaction of a path formula $\phi$ for a may- or must-execution $\rho = s_0 \lambda_{1} s_{1} \lambda_{2} \ldots$ of an MTS $\Mmm$
(we write $\rho_i=s_i$ to denote the $i$-th state of $\rho$, and $|\rho|$ to denote the number of states in $\rho$), denoted $[\Mmm,\rho \models^3 \phi]$, is
defined as ($\Mmm$ is omitted when clear from context):
\begin{description}
\item[(4)] $[\rho \models^3 \! (\Phi_1 \U \Phi_2)] \!=\! \left.\begin{cases} \ttv, & \!\!\text{if } \exists 0 \!\leq\! i \!\leq\! |\rho|. \big( [\rho_i \!\models^3 \! \Phi_2]\!=\!\ttv \land (\forall j \!<\! i.  [\rho_j \!\models^3 \Phi_1]\!=\!\ttv) \big) \\[1.1ex] \ffv, & \!\!\text{if } \begin{array}{@{} l @{}} \forall 0 \!\leq\! i \!\leq\! |\rho|. \big( \forall j \!<\! i. [\rho_j \!\models^3 \! \Phi_1] \!\neq\! \ffv \!\!\implies\!\!  [\rho_i \! \models^3 \! \Phi_2]\!=\!\ffv \big) \\ \land \ \forall i \!\geq\! 0. [\rho_i \!\models^3 \! \Phi_1] \!\neq\! \ffv \!\implies\!\! |\rho|=\infty\end{array} \\[1.1ex] \bot, & \!\!\text{otherwise} \end{cases}\right.$
\end{description}
A MTS $\Mmm$ satisfies a formula $\Phi$, written $[\Mmm \models^3 \Phi]=\ttv$, iff 
$\forall s_0 \in I. \, [s_0 \models^3 \Phi]=\ttv$. We say that $[\Mmm \models^3 \Phi]=\ffv$ if 
$\exists s_0 \in I. \, [s_0 \models^3 \Phi]=\ffv$. Otherwise, $[\Mmm \models^3 \Phi]=\bot$.

\begin{example} \label{exp:property}
Consider the FTS \vending\, and MTS $\joinasym(\vending)$ in Figures~\ref{fig:FTS} and \ref{fig:abstractFTS}. 
The property $\Phi_1=A (\neg \textit{r} \U \textit{r})$ states that in the initial state
along every execution will eventually reach the state where $\textit{r}$ holds.
Note that $[\vending \models \Phi_1]=\ffv$.
E.g., if the feature $\fc$  is enabled, a counter-example where the state $s_2$ that satisfies \textit{r} is never reached is:
$s_0 \to s_1 \to s_0 \to \ldots$. The set of violating products is
$\sbr{c}\!=\!\{ \{ \fc \},\linebreak[0] \{ \ff,\linebreak[0] \fc \} \} \subseteq \Kk^{VM}$.
However, $[\pi_{\sbr{\neg \fc}}(\vending) \models \Phi_1]=\ttv$.
We also have that $[\joinasym(\vending) \models^3 \Phi_1]=\bot$, since (1) there is a may-execution in $\joinasym(\vending)$ where $s_2$ is never reached:
$s_0 \to s_1 \to s_0 \to \ldots$, and (2) there is no must-execution that violates $\Phi_1$.

Consider the property $\Phi_2=E (\neg \textit{r} \U \textit{r})$, which describes a situation where
in the initial state
there exists an execution that will eventually reach $s_2$ that satisfies \textit{r}.
Note that $[\vending \models \Phi_2]=\ttv$, since even for variants with the feature $\fc$ there is a continuation from the state
$s_1$ to $s_2$. But, $[\joinasym(\vending) \models \Phi_2] = \bot$ since (1) there is no a must-execution in $\joinasym(\vending)$ that
reaches $s_2$ from $s_0$, and (2) there is a may-execution that satisfies $\Phi_2$.
\qed
\end{example}

\vspace{-1mm}
\section{Abstraction of FTSs}\label{sec:definition}

We now introduce the variability abstractions \cite{fase18} which
preserve full CTL. 
We start working with Galois connections\footnote{$\poset{L}{\leq_L} \galois{\alpha}{\gamma} \poset{M}{\leq_M}$ is a \emph{Galois connection} between complete lattices $L$ (concrete domain) and $M$ (abstract domain) iff $\alpha:L\to M$ and $\gamma:M \to L$ are total functions that satisfy: $\alpha(l) \leq_M m \iff l \leq_L \gamma(m)$, for all $l \in L, m \in M$.
 } between Boolean complete lattices of feature expressions, and then induce a notion of abstraction of FTSs.

The Boolean complete lattice of feature expressions (propositional formulae over $\Ff$) is: \((\FeatExp(\Ff)_{/\equiv},\models,\lor,\land,\true,\false,\neg)\).  The elements of the domain
$\FeatExp(\Ff)_{/\equiv}$ are equivalence classes of propositional formulae \(\psi\!\in\! \FeatExp(\Ff)\) obtained by quotienting by the semantic equivalence
$\equiv$. The ordering $\models$ is the standard entailment between propositional logics formulae,
whereas the least upper bound and the greatest lower bound are just logical disjunction and conjunction respectively.
Finally, the constant \false\ is the least, \true\ is the greatest element, and negation is the complement operator.

\paragraph{\textbf{Over-approximating abstractions.}}
The \emph{join abstraction}, $\joinasym$, replaces each feature expression $\psi$ with \true\ if there exists at least one configuration from $\Kk$ that satisfies $\psi$.
The abstract set of features is empty: $\joinasym(\Ff)=\emptyset$, and abstract set of configurations is a singleton: $\joinasym(\Kk) = \{ \true \}$.
The abstraction and concretization functions
between $\FeatExp(\Ff)$ and $\FeatExp(\emptyset)$ are:
\[
\joinasym(\psi) \!=\!
  \begin{cases}
    \true  & \textrm{if } \exists k \in \Kk. k \models \psi \\
    \false & \textrm{otherwise}
  \end{cases} \qquad \
\joingsym(\psi) \!=\!
  \begin{cases}
  \true & \text{if } \psi \text{ is } \true\\
  \bigvee_{k \in 2^\Ff\!\setminus\! \Kk} k & \text{if } \psi \text{ is } \false
  \end{cases}
\]
which form a Galois connection \cite{sttt16}.
In this way, we obtain a single abstract variant that includes all transitions occurring in any variant.
%
%



\paragraph{\textbf{Under-approximating abstractions.}}
The \emph{dual join abstraction}, $\widetilde{\joinasym}$, replaces each feature expression $\psi$ with \true\ if all configurations from $\Kk$ satisfy $\psi$.
The abstraction and concretization functions
between $\FeatExp(\Ff)$ and $\FeatExp(\emptyset)$, forming a Galois connection \cite{fase18}, are defined as \cite{DBLP:conf/sara/Cousot00}:
$\widetilde{\joinasym}=\neg \circ \joinasym \circ \neg$ and $\widetilde{\joingsym}=\neg \circ \joingsym \circ \neg$, that is:
\begin{equation*}
\widetilde{\joinasym}(\psi) =
\begin{cases}
    \true & \textrm{if } \forall k \in \Kk. k \models \psi \\
	\false  & \textrm{otherwise }
	
\end{cases}
\qquad
\widetilde{\joingsym}(\psi) \!=\!
  \begin{cases}
  \bigwedge_{k \in 2^\Ff \backslash \Kk} (\neg k) & \text{if } \psi \text{ is } \true\\
  \false & \text{if } \psi \text{ is } \false
  \end{cases}
\end{equation*}
In this way, we obtain a single abstract variant that includes only those transitions that occur in all variants.

\paragraph{\textbf{Abstract MTS and Preservation of CTL.}}
Given a Galois connection $(\joinasym,\joingsym)$ defined on the level of feature expressions,
we now define the abstraction of an FTS as an MTS with two transition relations: one (may)
preserving universal properties, and the other (must) preserving existential properties.
The may transitions describe the behaviour that is possible in some variant of the concrete FTS,
but not need be realized in the other
variants; whereas the must transitions describe behaviour that has to be present
in all variants of the FTS.
\begin{definition}
Given the FTS $\Fff=(S,Act,trans,I,AP,L,\Ff,\Kk,\delta)$, 
define MTS $\joinasym(\Fff)=(S,Act,trans^{\may},trans^{\must},I,AP,L)$
to be its \emph{abstraction},
where $trans^{\may} \!=\! \{ t \!\in\! trans \mid \joinasym(\delta(t)) \!=\!  \true \}$, and
 $trans^{\must} \!=\! \{ t \!\in\! trans \mid \widetilde{\joinasym}(\delta(t)) \!=\! \true \}$.
\end{definition}
Note that the abstract model $\joinasym(\Fff)$ has no variability in it,
i.e.\ it contains only one abstract configuration. 
We now show that 
the 3-valued semantics of the MTS $\joinasym(\Fff)$ is designed to be \emph{sound} in the sense that
it preserves both satisfaction ($\ttv$) and refutation ($\ffv$) of a formula from the abstract model
to the concrete one. However, if the truth value of a formula in the abstract model is $\bot$, then
its value over the concrete model is not known. 
We prove (see Appendix~\ref{app:proofs}):

\begin{theorem}[Preservation results] \label{theorem:sound}
For every $\Phi \in CTL$, we have:
\begin{description}
\item[(1)]
$[\joinasym(\Fff) \models^3 \Phi]\!=\!\ttv \, \implies \, [\Fff \models \Phi]\!=\!\ttv$.
\item[(2)]
$[\joinasym(\Fff) \models^3 \Phi]\!=\!\ffv \, \implies \, [\Fff \models \Phi]\!=\!\ffv$ and $[\pi_k(\Fff) \models \Phi]\!=\!\ffv$ for all $k \in \Kk$.
\end{description}
\end{theorem}

\paragraph{\textbf{Divide-and-conquer strategy.}}
The problem of evaluating $[\Fff \models \Phi]$ can be reduced to a number of smaller problems by
partitioning the configuration space $\Kk$.
Let the subsets $\Kk_1, \Kk_2, \ldots, \Kk_n$ form a \emph{partition} of the set $\Kk$.
Then, $[\Fff \models \Phi]=\ttv$ iff $[\pi_{\Kk_i}(\Fff) \models \Phi]=\ttv$ for all $i=1,\ldots,n$.
Also, $[\Fff \models \Phi]=\ffv$ iff $[\pi_{\Kk_j}(\Fff) \models \Phi]=\ffv$ for some $1 \leq j \leq n$.
By using Theorem~\ref{theorem:sound}, we obtain the following result.
\begin{corollary}
Let $\Kk_1, \Kk_2, \ldots, \Kk_n$ form a \emph{partition} of $\Kk$.
\begin{description}
\item[(1)]
If $[\joinasym(\pi_{\Kk_1}(\Fff)) \models \Phi]\!=\!\ttv \, \land \ldots \land \, [\joinasym(\pi_{\Kk_n}(\Fff)) \models \Phi]\!=\!\ttv$,
 then $[\Fff \models \Phi]\!=\!\ttv$.
 \item[(2)]
 If $[\joinasym(\pi_{\Kk_j}(\Fff)) \models \Phi]\!=\!\ffv$ for some $1 \!\leq\! j \leq n$,
 then $[\Fff \models \Phi]\!=\!\ffv$ and $[\pi_{k}(\Fff) \models \Phi]\!=\!\ffv$ for all $k \in \Kk_j$.
 \end{description}
\end{corollary}

\begin{example}
Recall the FTS $\vending$ of Fig.~\ref{fig:FTS}. 
Figure~\ref{fig:abstractFTS} shows the MTS $\joinasym({\vending})$, where the allowed (may) part of the behavior includes  the transitions that are associated with the optional features \fc\ and \ff\ in \vending, and the required (must) part 
includes transitions with the presence condition \true.
Consider the properties introduced in Example~\ref{exp:property}.
We have $[\joinasym(\vending) \models^3 \Phi_1]=\bot$ and $[\joinasym(\vending) \models^3 \Phi_2]=\bot$,
so we cannot conclude whether $\Phi_1$ and $\Phi_2$
are satisfied by \vending\, or not.
\qed
\end{example} 

\vspace{-1mm}
\section{Game-based Abstract Lifted Model Checking}

The 3-valued model checking game \cite{DBLP:journals/tocl/ShohamG07,DBLP:journals/iandc/ShohamG10}
on an MTS $\Mmm$ with state set $S$, a state $s \in S$, and
a CTL formula $\Phi$ is played by Player $\forall$ and Player $\exists$ in order to
evaluate $\Phi$ in $s$ of $\Mmm$.
The goal of Player $\forall$ is either to refute $\Phi$ on $\Mmm$ or to
prevent Player $\exists$ from verifying it.
The goal of Player $\exists$ is either to verify $\Phi$ on $\Mmm$ or to
prevent Player $\forall$ from refuting it.
The \emph{game board} is the Cartesian product $S \times sub(\Phi)$, where $sub(\Phi)$ is defined as:
\[
\begin{array}{l}
\text{if } \Phi \!=\! \true, \false, l, \text{then } sub(\Phi) \!=\! \{ \Phi \}; \, \text{if } \Phi \!=\! \text{\AE}\! \bigcirc\! \Phi_1, \text{then } sub(\Phi) \!=\! \{ \Phi \} \!\cup\! sub(\Phi_1) \\
\text{if } \Phi = \Phi_1 \land \Phi_2, \Phi_1 \lor \Phi_2, \text{ then } sub(\Phi) = \{ \Phi \} \cup sub(\Phi_1) \cup sub(\Phi_2) \\
\text{if } \Phi = \text{\AE} (\Phi_1 \U \Phi_2),  \text{\AE} (\Phi_1 \V \Phi_2), \text{ then } sub(\Phi) = exp(\Phi) \cup sub(\Phi_1) \cup sub(\Phi_2)
\end{array}
\]
where \AE\ ranges over both $A$ and $E$.
The expansion  $exp(\Phi)$ is defined as:
\[
\begin{array}{l}
\Phi = \text{\AE} (\Phi_1 \U \Phi_2): exp(\Phi) = \{ \Phi, \Phi_2 \lor (\Phi_1 \land \text{\AE} \bigcirc \Phi), \Phi_1 \land \text{\AE} \bigcirc \Phi, \text{\AE} \bigcirc \Phi \} \\
\Phi = \text{\AE} (\Phi_1 \V \Phi_2): exp(\Phi) = \{ \Phi, \Phi_2 \land (\Phi_1 \lor \text{\AE} \bigcirc \Phi), \Phi_1 \lor \text{\AE} \bigcirc \Phi, \text{\AE} \bigcirc \Phi \}
\end{array}
\]

A  \emph{single play} from $(s,\Phi)$ is a possibly infinite sequence of configurations $C_0 \to_{p_0} C_1 \to_{p_1} C_2 \to_{p_2} \ldots$,
where $C_0=(s,\Phi)$, $C_i \in S \times sub(\Phi)$, and $p_i \in \{ \text{Player } \forall, \text{Player } \exists \}$.
The subformula in $C_i$ determines which player $p_i$ makes the next move.
The possible moves at each configuration are:
\begin{description}
\item[(1)] $C_i=(s,\false)$, $C_i=(s,\true)$, $C_i=(s,l)$: the play is finished.
Such configurations are called \emph{terminal}.
\item[(2)] if $C_i\!=\!(s,A \bigcirc \Phi)$, Player $\forall$ chooses a must-transition $s \,\TR{}\, s'$ (for refutation)
or a may-transition $s \,\TR{}\, s'$ of $\Mmm$ (to prevent satisfaction), and $C_{i+1}\!=\!(s',\Phi)$.
\item[(3)] if $C_i\!=\!(s,E \bigcirc \Phi)$, Player $\exists$ chooses a must-transition $s \,\TR{}\, s'$ (for satisfaction)
or a may-transition $s \,\TR{}\, s'$ of $\Mmm$ (to prevent refutation), and $C_{i+1}\!=\!(s',\Phi)$.
\item[(4)] if $C_i=(s,\Phi_1 \land \Phi_2)$, then Player $\forall$ chooses $j \in \{1,2\}$ and $C_{i+1}=(s,\Phi_j)$.
\item[(5)] if $C_i=(s,\Phi_1 \lor \Phi_2)$, then Player $\exists$ chooses $j \in \{1,2\}$ and $C_{i+1}=(s,\Phi_j)$.
\item[(6),(7)] if $C_i=(s,\text{\AE} (\Phi_1 \U \Phi_2))$, then $C_{i+1}=(s,\Phi_2 \lor (\Phi_1 \land \text{\AE} \bigcirc \text{\AE} (\Phi_1 \U \Phi_2)))$.
\item[(8),(9)] if $C_i=(s,\text{\AE} (\Phi_1 \V \Phi_2))$, then $C_{i+1}=(s,\Phi_2 \land (\Phi_1 \lor \text{\AE} \bigcirc \text{\AE} (\Phi_1 \V \Phi_2)))$.
\end{description}
The moves $(6)-(9)$ are deterministic, thus any player can make them.

A play is a \emph{maximal play} iff it is infinite or ends in a terminal configuration. A play is infinite \cite{DBLP:series/txcs/Stirling01} iff
there is exactly one subformula of the form $A\U$, $A\V$, $E\U$, or $E\V$ that occurs infinitely often in the play.
Such a subformula is called a \emph{witness}. We have the following \emph{winning criteria}:
\begin{itemize}
\item Player $\forall$ \emph{wins} a (maximal) play iff in each configuration of the form $C_i=(s,A \bigcirc \Phi)$, Player $\forall$
chooses a move based on must-transitions and one of the following holds: (1) the play is finite and ends in
a terminal configuration of the form $C_i=(s,\false)$ or $C_i=(s,a)$ where $a \not\in L(s)$ or $C_i=(s,\neg a)$ where $a \in L(s)$; (2) the play is infinite and
the witness is of the form $A\U$ or $E\U$.
\item Player $\exists$ \emph{wins} a (maximal) play iff in each configuration of the form $C_i=(s,E \bigcirc \Phi)$, Player $\exists$
chooses a move based on must-transitions and one of the following holds: (1) the play is finite and ends in
a terminal configuration of the form $C_i=(s,\true)$ or $C_i=(s,a)$ where $a \in L(s)$ or $C_i=(s,\neg a)$ where $a \not\in L(s)$; (2) the play is infinite and
the witness is of the form $A\V$ or $E\V$.
\item Otherwise, the play ends in a \emph{tie}.
\end{itemize}

A \emph{strategy} is a set of rules for a player, telling the player which move to choose in the current configuration.
A \emph{winning strategy} from $(s,\Phi)$ is a set of rules allowing the player to win every play that starts at $(s,\Phi)$
if he plays by the rules.
It was shown in \cite{DBLP:journals/tocl/ShohamG07,DBLP:journals/iandc/ShohamG10} that the model checking problem
of evaluating $[\Mmm, s \models^3 \Phi]$  can be reduced to the problem
of finding which player has a winning strategy from $(s,\Phi)$ (i.e.\ to solving
the given 3-valued model checking game).

The algorithm proposed in \cite{DBLP:journals/tocl/ShohamG07,DBLP:journals/iandc/ShohamG10} for
solving the given 3-valued model checking game consists of two parts.
First, it constructs a \emph{game-graph}, then it runs an \emph{algorithm for coloring} the game-graph.
The 
game-graph is $G_{\Mmm \times \Phi} = (N,E)$ where $N \subseteq S \times sub(\Phi)$
is the set of nodes and $E \subseteq N \times N$ is the set of edges. $N$ contains a node
for each configuration that was reached during the construction of the game-graph that starts from
initial configurations $I \times \{\Phi\}$ in a BFS manner, and $E$ contains an edge for
each possible move that was applied. The nodes of the game-graph can be classified as: terminal
nodes, $\land$-nodes, $\lor$-nodes, $A\bigcirc$-nodes, and $E\bigcirc$-nodes.
Similarly, the edges can be classified as: progress edges, which originate in
$A\bigcirc$ or $E\bigcirc$ nodes and reflect real transitions of the MTS $\Mmm$, and auxiliary nodes,
which are all other edges.
We distinguish two types of progress edges, two types of children, and two types of SCCs
(Strongly Connected Components).
\emph{Must-edges} (\emph{may-edges}) are edges based on must-transitions (may-transitions) of MTSs.
A node $n'$ is a \emph{must-child} (\emph{may-child}) of the node $n$ if there exists a must-edge (may-edge) $(n,n')$.
A \emph{must-SCC} (\emph{may-SCC}) 
is an SCC in which all progress edges are must-edges (may-edges).

The game-graph is partitioned into its may-Maximal SCCs
(may-MSCCs), denoted $Q_i$'s. This partition induces a partial order $\leq$ on the $Q_i$'s, such that
edges go out of a set $Q_i$ only to itself or to a smaller set $Q_j$.
The partial order is extended to a total order $\leq$ arbitrarily.
The \emph{coloring algorithm} processes the $Q_i$'s according to $\leq$, bottom-up.
Let $Q_i$ be the smallest set that is not fully colored.
The nodes of $Q_i$ are colored in two phases, as follows.

\noindent
\emph{Phase 1.} Apply these rules to all nodes in $Q_i$ until none of them is applicable.
\begin{itemize}
\item A terminal node $C$ is colored: by $T$ if Player $\exists$ wins in it (when $C=(s,\true)$ or $C=(s,a)$ with $a \in L(s)$ or $C=(s,\neg a)$ with $a \not\in L(s)$);
and by $F$ if Player $\forall$ wins in it (when $C=(s,\false)$ or $C=(s,a)$ with $a \not\in L(s)$ or $C=(s,\neg a)$ with $a \in L(s)$).
\item An $A \bigcirc$ node is colored: by $T$ if all its may-children are colored by $T$; by $F$ if
it has a must-child colored by $F$; by $?$ if all its must-children are colored by $T$ or $?$, and it has
a may-child colored by $F$ or $?$.
\item An $E \bigcirc$ node is colored: by $T$ if it has a must-child colored by $T$; by $F$ if
all its may-children are colored by $F$; by $?$ if it has a may-child colored by $T$ or $?$, and all its must-children are colored by $F$ or $?$.
\item An $\land$-node ($\lor$-node) is colored: by $T$ ($F$) if both its children are colored by $T$ ($F$);
by $F$ ($T$) if it has a child that is colored by $F$ ($T$); by $?$ if it has a child colored by $?$ and
the other child is colored by $?$ or $T$ ($F$).
\end{itemize}

\noindent
\emph{Phase 2.}
If after propagation of the rules of Phase 1, there are still nodes in $Q_i$ that remain uncolored,
then $Q_i$ must be a non-trivial may-MSCC that has exactly one witness. We consider two cases.
\begin{description}
\item[Case \U.] The witness is of the form $A(\Phi_1 \U \Phi_2)$ or $E(\Phi_1 \U \Phi_2)$. \\
\noindent
\emph{Phase 2a.} Repeatedly color by $?$ each node in $Q_i$ that satisfies one of the following conditions,
until there is no change: \\
(1) An $A \bigcirc$ node that all its must-children are colored by $T$ or $?$;
(2) An $E \bigcirc$ node that has a may-child colored by $T$ or $?$;
(3) An $\land$ node that both its children are colored $T$ or $?$;
(4) An $\lor$ node that has a child colored by $T$ or $?$. \\
In fact, each node for which the $F$ option is no longer possible according
to the rules of Phase 1 is colored by $?$. \\
\emph{Phase 2b.} Color the remaining nodes in $Q_i$ by $F$.

\item[Case \V.] The witness is of the form $A(\Phi_1 \V \Phi_2)$ or $E(\Phi_1 \V \Phi_2)$. \\
\noindent
\emph{Phase 2a.} Repeatedly color by $?$ each node in $Q_i$ that satisfies one of the following conditions,
until there is no change. \\
(1) An $A \bigcirc$ node that has a may-child colored by $F$ or $?$;
(2) An $E \bigcirc$ node that all its must-children are colored by $F$ or $?$;
(3) An $\land$ node that has a child colored $F$ or $?$;
(4) An $\lor$ node that both its children are colored $F$ or $?$. \\
In fact, each node for which the $T$ option is no longer possible according
to the rules of Phase 1 is colored by $?$. \\
\emph{Phase 2b.} Color the remaining nodes in $Q_i$ by $T$.
\end{description}
The result of the coloring is a \emph{3-valued coloring function} $\chi : N \to \{ T, F, ? \}$.

\tikzset{
    standard/.style={draw=black,->,>=latex,thick,solid},
    dash/.style={draw,->,>=latex,thick,dashed}
}
\tikzset{
red/.style={shape=rectangle, rounded corners,draw=black, dashed, align=center,top color=white, bottom color=red},
blue/.style={shape=rectangle, rounded corners,draw=black, solid, align=center,top color=white, bottom color=green},
white/.style={shape=rectangle, rounded corners,draw=black, solid, align=center,top color=white, bottom color=white}
}

\begin{figure}[t]
\centering
\centering
\hspace{-2cm}
\begin{tikzpicture}[->,>=stealth',level distance=2.9em,sibling distance=5.75em,level 1/.style={sibling distance=5.75em},
    level 2/.style={sibling distance=6.2em},level 2/.style={sibling distance=6.4em}]
  \node[white](0) {\tiny $(s_0,A(\neg \textit{r} \U \textit{r}))$}
    child { node[white] {\tiny $(s_0,\textit{r} \lor (\neg \textit{r} \land A \bigcirc A (\neg \textit{r} \U \textit{r})))$} edge from parent[standard]
      child { node[red,label=left:{\tiny $Q1$}] {\tiny $(s_0, \sr)$} edge from parent[standard] }
      child { node[white] {\tiny $(s_0,\neg \sr \land A \bigcirc A (\neg \sr \U \sr))$} edge from parent[standard]
        child { node[blue,label=left:{\tiny $Q2$}] {\tiny $(s_0,\neg \sr)$} edge from parent[standard] }
        child { node[white] {\tiny $(s_0,A \bigcirc A (\neg \sr \U \sr))$} edge from parent[standard]
                child { node[white,label=above:{\textit{pay}}] {\tiny $(s_1,A (\neg \sr \U \sr))$} edge from parent[dash]
                        child { node[white] {\tiny $(s_1,\sr \lor (\neg \sr \land A \bigcirc A (\neg \sr \U \sr)))$} edge from parent[standard]
                                   child { node[red,label=left:{\tiny $Q3$}] {\tiny $(s_1, \sr)$} edge from parent[standard] }
                                   child { node[white] {\tiny $(s_1,\neg \sr \land A \bigcirc A (\neg \sr \U \sr))$} edge from parent[standard]
                                        child { node[blue,label=above:{\tiny $Q4$}] {\tiny $(s_1,\neg \sr)$} edge from parent[standard] }
                                        child  { node[white,label=below:{\tiny failure node}] {\tiny $(s_1,A \bigcirc A (\neg \sr \U \sr))$} edge from parent[standard] }
                                   }
                        }
                        child {edge from parent[draw=none]}
                }
                child { node[blue,label=above:{\textit{free}}] {\tiny $(s_2,A (\neg \sr \U \sr))$} edge from parent[dash]
                        child {edge from parent[draw=none]}
                        child { node[blue] {\tiny $(s_2,\sr \lor (\neg \sr \land A \bigcirc A (\neg \sr \U \sr)))$} edge from parent[standard]
                                   child { node[blue,label=above:{\tiny $Q5$}] {\tiny $(s_2, \sr)$} edge from parent[standard] }
                                   child { node[red] {\tiny $(s_2,\neg \sr \land A \bigcirc A (\neg \sr \U \sr))$} edge from parent[standard]
                                        child { node[red,label=above:{\tiny $Q6$}] {\tiny $(s_2,\neg \sr)$} edge from parent[standard] }
                                        child { node[white] {\tiny $(s_2,A \bigcirc A (\neg \sr \U \sr))$} edge from parent[standard] }
                                   }
                        }
                }
            } }
                                     };

       \path[draw, dashed]
            (0-1-2-2-1-1-2-2) edge [relative, out=90, in=80, looseness=1.491] node[right] {$\textit{cancel}$} (0);
       \path[draw]
            (0-1-2-2-1-1-2-2) edge  node[left] {$\textit{drink}$} (0-1-2-2-2)
            (0-1-2-2-2-2-2-2) edge [bend right=70] node[left] {$\textit{take}$} (0);
\end{tikzpicture}
\vspace{-9.77mm}
\caption{The colored game-graph for $\joinasym(\vending)$ and $\Phi_1=A(\neg \sr \U \sr)$.}
\label{fig:refine1}
\end{figure}

\begin{theorem}[\cite{DBLP:journals/tocl/ShohamG07}]\label{theorem:color}
For each $n = (s,\Phi') \in G_{\Mmm \times \Phi}$:
\begin{description}
\item[(1)] $[(\Mmm,s) \models^3 \Phi']=\ttv$ iff $\chi(n)=T$ iff Player $\exists$ has a winning strategy at $n$.
\item[(2)] $[(\Mmm,s) \models^3 \Phi']=\ffv$ iff $\chi(n)=F$ iff Player $\forall$ has a winning strategy at $n$.
\item[(3)] $[(\Mmm,s) \!\models^3\! \Phi']\!=\!\!\bot$ iff $\chi(n)\!=?$ iff none of players has a winning strategy at $n$.
\end{description}
\end{theorem}
Using Theorem~\ref{theorem:sound} and Theorem~\ref{theorem:color},
given the colored game-graph of the MTS $\joinasym(\Fff)$, if all its initial nodes are colored by $T$
then $[\Fff \models \Phi] = \ttv$, if at least one of them is colored by $F$ then $[\Fff \models \Phi] = \ffv$.
Otherwise, we do not know.

\begin{example}
The colored game-graph for the MTS $\joinasym({\vending})$ and $\Phi_1=A (\neg \textit{r} \U \textit{r})$
is shown in Fig.~\ref{fig:refine1}. Green, red (with dashed borders), and white nodes denote nodes colored by $T$, $F$, and $?$, respectively.
The partitions from $Q_1$ to $Q_6$ consist of a single node shown in Fig.~\ref{fig:refine1}, while
$Q_7$ contains all the other nodes.
The initial node $(s_0,\Phi_1)$ is colored by $?$, so
we obtain an indefinite answer. 
\qed
\end{example}

\vspace{-1.5mm}
\section{Incremental Refinement Framework}

Given an FTS $\pi_{\Kk'}(\Fff)$ with a configuration set $\Kk' \subseteq \Kk$,
we show how to exploit the game-graph of the abstract MTS $\Mmm = \joinasym(\pi_{\Kk'}(\Fff))$
in order to
do refinement in case that the model checking resulted in an indefinite answer.
The refinement consists of two parts. First, we use the information gained by the
 coloring algorithm of $G_{\Mmm \times \Phi}$ in order to split the single abstract configuration
 $\true \in \joinasym(\Kk')$ that represents the whole concrete configuration set $\Kk'$.
We then construct the refined abstract models, using the refined abstract configurations.

There are a failure node and a failure reason associated with an indefinite answer.
The goal in the refinement is to find and eliminate at least one of the failure reasons.
\begin{definition}
A node $n$ is a \emph{failure node} if it is colored by $?$, whereas none of its children was colored
by $?$ at the time $n$ got colored by the coloring algorithm.
\end{definition}
Such failure node can be seen as the point where the loss of information occurred, so
we can use it in the refinement step to change the final model checking result. 

\begin{figure}[t]
\fbox{
\begin{minipage}{88ex}
\textbf{Algorithm.}  \texttt{Verify($\Fff,\Kk,\Phi$)}

\begin{itemize}
\item[1]
Check by game-based model checking algorithm [$\joinasym(\Fff) \models^3 \Phi$]?
\item[2]
If the result is $\ttv$, then return that $\Phi$ is satisfied for all variants in $\Kk$.
If the result is $\ffv$, then return that $\Phi$ is violated for all variants in $\Kk$.
\item[3]
Otherwise, an indefinite result is returned.
Let the may-edge from $n=(s,\Phi_1)$ to $n'=(s',\Phi_1')$ be the reason for failure, and let
$\psi$ be the feature expression guarding the transition from $s$ to $s'$ in $\Fff$.
We generate $\Fff_1=\pi_{\sbr{\psi}}(\Fff)$ and $\Fff_2=\pi_{\sbr{\neg \psi}}(\Fff)$,
and call \texttt{Verify($\Fff_1,\Kk \cap \sbr{\psi},\Phi$)} and \texttt{Verify($\Fff_2,\Kk \cap \sbr{\neg \psi},\Phi$)}.
\end{itemize}
\end{minipage}
}
\vspace{-2.5mm}
\caption{The  Refinement Procedure that checks $[\Fff \models \Phi]$. 
} \label{fig:ARP}
\end{figure}

\begin{lemma}[\cite{DBLP:journals/tocl/ShohamG07}]\label{lemma:reason}
A failure node is one of the following.
\begin{itemize}
\item An $A \bigcirc$-node ($E \bigcirc$-node) that has a may-child colored by $F$ ($T$).
\item An $A \bigcirc$-node ($E \bigcirc$-node) that was colored during Phase 2a based on an
$A\U$ ($A\V$) witness, and has a may-child colored by $?$.
\end{itemize}
\end{lemma}
Given a failure node $n=(s,\Phi)$,
suppose that its may-child is $n'=(s',\Phi_1')$ as identified in Lemma~\ref{lemma:reason}.
Then the may-edge from $n$ to $n'$ is considered as \emph{the failure reason}.
Since the failure reason is a may-transition in the abstract MTS $\joinasym(\pi_{\Kk'}(\Fff))$, it needs to be refined in order
to result either in a must transition or no transition at all.
Let $s \TR{\alpha/\!\psi} s'$ be the transition in the concrete model $\pi_{\Kk'}(\Fff)$
corresponding to the above (failure) may-transition.
We split the configuration space $\Kk'$ into $\sbr{\psi}$ and $\sbr{\neg \psi}$ subsets,
and we partition $\pi_{\Kk'}(\Fff)$ in $\pi_{\sbr{\psi} \cap \Kk'}(\Fff)$ and
$\pi_{\sbr{\neg \psi} \cap \Kk'}(\Fff)$. Then, we repeat the verification process based on abstract models
$\joinasym(\pi_{\sbr{\psi} \cap \Kk'}(\Fff))$ and $\joinasym(\pi_{\sbr{\neg \psi} \cap \Kk'}(\Fff))$.
Note that, in the former, $\joinasym(\pi_{\sbr{\psi} \cap \Kk'}(\Fff))$, $s \TR{\alpha} s'$ becomes a must-transition,
while in the latter, $\joinasym(\pi_{\sbr{\neg \psi} \cap \Kk'}(\Fff))$, $s \TR{\alpha} s'$ is removed.
The complete refinement procedure is shown in Fig.~\ref{fig:ARP}.
We prove that (see Appendix~\ref{app:proofs}):
\begin{theorem} \label{theorem:correct}
The procedure \texttt{Verify($\Fff,\Kk,\Phi$)} terminates and is correct.
\end{theorem}

\begin{example}
We can do a failure analysis on the game-graph of $\joinasym(\vending)$ in Fig.~\ref{fig:refine1}.
The failure node is $(s_1,A \bigcirc A (\neg \textit{r} \U \textit{r}))$ and the reason is the may-edge
$(s_1,A \bigcirc A (\neg \textit{r} \U \textit{r})) \TR{\textit{cancel}} (s_0, A (\neg \textit{r} \U \textit{r}))$.
The corresponding concrete transition in \vending\ is $s_1 \TR{\textit{cancel}/\!\fc} s_0$.
So, we partition the configuration space $\displaystyle\Kk^{\textsc{VM}}$ into subsets $\sbr{\fc}$ and $\sbr{\neg \fc}$, and in the next second
iteration we consider FTSs $\pi_{\sbr{\fc}}(\vending)$ and $\pi_{\sbr{\neg \fc}}(\vending)$. \qed
\end{example}

The game-based model checking algorithm provides us with a convenient framework to use
 results from previous iterations and avoid unnecessary calculations. At the end of the $i$-th iteration
of abstraction-refinement, we remember those nodes that were colored by
definite colors. Let $D$ denote the set of such nodes.
Let $\chi_D: D \to \{ T, F \}$ be the coloring function that maps each node in $D$
to its definite color.
The incremental approach uses this information both in the construction of the game-graph
and its coloring.
During the construction of a new refined game-graph performed in a BFS manner in the next $i+1$-th iteration, we prune the game-graph
in nodes that are from $D$. When a node $n \in D$ is encountered, we add $n$ to the game-graph and
do not continue to construct the game-graph from $n$ onwards.
That is, $n \in D$ is considered as terminal node and colored by its
previous color.
As a result of this pruning, only the reachable sub-graph that was previously colored by $?$ is refined.

\begin{figure}[t]
\centering
\begin{minipage}[b]{0.65\textwidth}
\hspace{-2cm}
\centering
\begin{tikzpicture}[->,>=stealth',level distance=2.83em,sibling distance=5.75em,level 1/.style={sibling distance=5.75em},
    level 2/.style={sibling distance=6.2em},level 2/.style={sibling distance=6.4em}]
  \node[white](0) {\tiny $(s_0,A(\neg \sr \U \sr))$}
    child { node[white] {\tiny $(s_0,\sr \lor (\neg \sr \land A \bigcirc A (\neg \sr \U \sr)))$} edge from parent[standard]
      child { node[red,label=left:{\tiny $Q1$}] {\tiny $(s_0, \sr)$} edge from parent[standard] }
      child { node[white] {\tiny $(s_0,\neg \sr \land A \bigcirc A (\neg \sr \U \sr))$} edge from parent[standard]
        child { node[blue,label=left:{\tiny $Q2$}] {\tiny $(s_0,\neg \sr)$} edge from parent[standard] }
        child { node[white,label=below:{\tiny failure node}] {\tiny $(s_0,A \bigcirc A (\neg \sr \U \sr))$} edge from parent[standard]
                child { node[white,label=above:{\textit{pay}}] {\tiny $(s_1,A (\neg \sr \U \sr))$} edge from parent[dash]
                        child { node[white] {\tiny $(s_1,\sr \lor (\neg \sr \land A \bigcirc A (\neg \sr \U \sr)))$} edge from parent[standard]
                                   child { node[red,label=above:{\tiny $Q3$}] {\tiny $(s_1, \sr)$} edge from parent[standard] }
                                   child { node[white] {\tiny $(s_1,\neg \sr \land A \bigcirc A (\neg \sr \U \sr))$} edge from parent[standard]
                                        child { node[blue,label=above:{\tiny $Q4$}] {\tiny $(s_1,\neg \sr)$} edge from parent[standard] }
                                        child  { node[white] {\tiny $(s_1,A \bigcirc A (\neg \sr \U \sr))$} edge from parent[standard] }
                                   }
                        }
                        child {edge from parent[draw=none]}
                }
                child { node[blue,label=above:{\textit{free}}] {\tiny $(s_2,A (\neg \sr \U \sr))$} edge from parent[dash]
                }
            } }
                                     };


       \path[draw]
            (0-1-2-2-1-1-2-2) edge [relative, out=90, in=80, looseness=1.491] node[right] {$\textit{cancel}$} (0);
       \path[draw]
            (0-1-2-2-1-1-2-2) edge [bend right=45] node[left] {$\textit{drink}$} (0-1-2-2-2);
\end{tikzpicture}
\vspace{-9.78mm}
\caption{$G_{\joinasym(\pi_{\sbr{\fc}}(\vending)) \times \Phi_1}$.}
\label{fig:refine2-2}
\end{minipage}%
\begin{minipage}[b]{.33\textwidth}
\centering
\begin{tikzpicture}[->,>=stealth',shorten >=0.9pt,auto,node distance=1.4cm, semithick]
  \tikzstyle{every state}=[minimum size=.2pt,initial text={{}}]

  \node[initial above,state] (A)     {\tiny $s_0$};
  \node[state]         (B) [right of=A] {\tiny $s_1$};
  \node[state]         (C) [right of=B] {\tiny $s_2$};

  \path[font=\tiny]
        (B) edge node {$\textit{drink}$} (C)
        (C) edge [bend left] node {$\textit{take}$} (A)
        (B) edge [bend right=45] node[above,sloped] {$\textit{cancel}$} (A);

  \path[draw, dashed, font=\tiny]
  (A) edge node {$\textit{pay}$} (B)
  (A) edge [bend left=75]  node[above,sloped] {{$\textit{free}$}} (C);

\end{tikzpicture}
\vspace{-2.55mm}
\caption{\scriptsize $\joinasym(\pi_{\sbr{\fc}}(\vending))$}
\label{fig:abstractFTSc}
\end{minipage}%
\end{figure}

\begin{example}
The property $\Phi_1$ holds for $\pi_{\sbr{\neg \fc}}(\vending)$.
We show the model $\joinasym(\pi_{\sbr{\neg \fc}}(\vending))$ and
the game-graph $G_{\joinasym(\pi_{\sbr{\neg \fc}}(\vending)) \times \Phi_1}$ 
in Fig.~\ref{fig:abstractFTSnc} and Fig.~\ref{fig:refine2-1} (see Appendix~\ref{app:figures}),
where the initial node is colored by $T$.
On the other hand, we obtain an indefinite answer for $\pi_{\sbr{ \fc}}(\vending)$.
The model  $\joinasym(\pi_{\sbr{ \fc}}(\vending))$ is shown in Fig.~\ref{fig:abstractFTSc}, whereas the final colored game-graph $G_{\joinasym(\pi_{\sbr{\fc}}(\vending)) \times \Phi_1}$ is given in Fig.~\ref{fig:refine2-2}.
The failure node is $(s_0,A \bigcirc A (\neg \textit{r} \U \textit{r}))$, and the reason is the may-edge
$(s_0,A \bigcirc A (\neg \textit{r} \U \textit{r})) \to (s_1, A (\neg \textit{r} \U \textit{r}))$.
The corresponding concrete transition in $\pi_{\sbr{ \fc}}(\vending)$ is $s_0 \TR{\textit{pay}/\!\neg \ff} s_1$.
So, in the next third iteration we consider FTSs $\pi_{\sbr{\fc \land \neg \ff}}(\vending)$ and $\pi_{\sbr{\fc \land \ff}}(\vending)$.

The graph $G_{\joinasym(\pi_{\sbr{\fc \land \neg \ff}}(\vending)) \times \Phi_1}$ is shown in Fig.~\ref{fig:refine3-1} (see Appendix~\ref{app:figures}),
where the initial node is colored by $F$ in Phase 2b.
$G_{\joinasym(\pi_{\sbr{\fc \land \ff}}(\vending)) \times \Phi_1}$ is shown in Fig.~\ref{fig:refine3-2} (see Appendix~\ref{app:figures}),
where the initial node is colored by $T$.

In the end, we conclude that $\Phi_1$ is satisfied by the variants $\{ \neg \fc \land \neg \ff, \neg \fc \land \ff, \fc \land \ff \}$,
and $\Phi$ is violated by the variant $\{ \fc \land \neg \ff \}$.

On the other hand, we need two iterations to conclude that $\Phi_2=E (\neg \textit{r} \U \textit{r})$
is satisfied by all variants in $\displaystyle\Kk^{\textsc{VM}}$ (see Appendix~\ref{app:figures2} for details).
\qed
\end{example}

\vspace{-1.5mm}

\section{Evaluation}

To evaluate our approach, we use a synthetic example to demonstrate specific
characteristics of our approach, and the \elevator\, model which is often
used as benchmark in SPL community \cite{DBLP:journals/scp/PlathR01,model-checking-spls,sttt16,fase18,programming17}.
We compare (1) our abstraction-refinement procedure \texttt{Verify} with
the game-based model checking algorithm implemented in Java from scratch vs.
(2) family-based version of the \smv model checker,
denoted \fsmv, which
implements the standard lifted model checking algorithm \cite{classen-model-checking-spls-icse11}.
For each experiment, we measure \textsc{T(ime)} to perform an analysis task, and \textsc{Call} which is the number of
times an approach calls the model checking engine. 
All experiments
were executed on a 
64-bit Intel$^\circledR$Core$^{TM}$ i5-3337U CPU running at 1.80 GHz with 8 GB
memory. All experimental data is available from: \url{https://aleksdimovski.github.io/automatic-ctl.html}.

\vspace{-1mm}
\paragraph{Synthetic example.}
The FTS $M_n$ (where $n>0$) consists of $n$ features $A_1, \ldots, A_n$
and an integer data variable $x$, such that the set $AP$ consists of all
evaluations of $x$ which assign nonnegative integer values to $x$.
The set of valid configurations is $\Kk_{n} = 2^{\{A_1, \ldots, A_n\}}$.
$M_n$ has a tree-like structure, where in the root is the initial
state with $x=0$. In each level $k$ ($k \geq 1$), there are two states
that can be reached with two transitions leading from a state from a previous level.
One transition is allowable for variants with the feature $A_k$ enabled, so that in
the target state the variable's value is $x+2^{k-1}$ where $x$ is its value in the source state, whereas
the other transition is allowable for variants with $A_k$ disabled,
so that the value of $x$ does not change. 
For example, $M_2$ is shown in Fig.~\ref{fig:FTS-M2}, where in each state we show the current value of $x$
and all transitions have the silent action $\tau$.

\tikzset{
circle/.style={shape=circle, draw=black, solid, align=center}
}

\begin{figure}[t]
\centering
\begin{minipage}[b]{.33\textwidth}
\centering
\begin{tikzpicture}[->,>=stealth',shorten >=0.9pt,auto,node distance=1.1cm, semithick]
  \tikzstyle{every state}=[minimum size=.2pt,initial text={{}}]

  \node[initial left,state] (A)     []  {\tiny $0$};
  \node[state]         (B) [above right of=A] {\tiny $1$};
  \node[state]         (C) [below right of=A] {\tiny $0$};
  \node[state]         (D) [above right of=B] {\tiny $3$};
  \node[state]         (E) [right of=B] {\tiny $1$};
  \node[state]         (F) [above right of=C] {\tiny $2$};
  \node[state]         (G) [right of=C] {\tiny $0$};

  \path[font=\tiny] (A) edge node[above, sloped] {$\tau /\! A_1$} (B)
                    (A) edge node[below, sloped] {$\tau /\!\neg A_1$} (C)
                    (B) edge node[above, sloped] {$\tau /\! A_2$} (D)
                    (B) edge node[below, sloped] {$\tau /\!\neg A_2$} (E)
                    (C) edge node[above, sloped] {$\tau /\! A_2$} (F)
                    (C) edge node[below, sloped] {$\tau /\!\neg A_2$} (G)
                    (D) edge [in=60, out=30, looseness=9] node[above,pos=.9] {$\tau$} (D)
                    (E) edge [in=60, out=30, looseness=9] node[above,pos=.9] {$\tau$} (E)
                    (F) edge [in=60, out=30, looseness=9] node[below,pos=.2] {$\tau$} (F)
                    (G) edge [in=60, out=30, looseness=9] node[above,pos=.9] {$\tau$} (G);

\end{tikzpicture}
\vspace{-2.5mm}
\caption{\small The model $M_2$.}
\vspace{-1mm}
\label{fig:FTS-M2}
\end{minipage}%
\begin{minipage}[b]{.65\textwidth}
\centering
\begin{tabular}{||@{}c@{}||c|c|c|c||c|c|c|c||}  \hhline{|t:=:t:=:=:=:=:t:=:=:=:=:t|}
 & \multicolumn{4}{c||}{$\Phi$}     & \multicolumn{4}{c||}{$\Phi'$} \\ \cline{2-9} 
 ~$n$~             & \multicolumn{2}{c|}{\small{\fsmv}} & \multicolumn{2}{c||}{~\small{\texttt{Verify}}~}  & \multicolumn{2}{c|}{\small{\fsmv}} & \multicolumn{2}{c||}{~\small{\texttt{Verify}}~} \\ \cline{2-9}
 & \textsc{Call} & \textsc{T} & \textsc{Call} & \textsc{T} & \textsc{Call} & \textsc{T} & \textsc{Call} & \textsc{T} \\
 \hhline{|:=::=:=:=:=::=:=:=:=:|}
2 & 1 & 0.08  & 1 & 0.07 &  1  & 0.08  & 5  & 0.83    \\ \hhline{||-||-|-|-|-||-|-|-|-||}
7 & 1 & 1.64  & 1 & 0.16 &  1  & 1.68  & 15 & 2.68 \\ \hhline{||-||-|-|-|-||-|-|-|-||}
10 & 1 & 992.80 & 1 & 0.68 &  1  & 1019.27  & 21 & 4.57  \\ \hhline{||-||-|-|-|-||-|-|-|-||}
11 & 1 & \tiny{infeasible} & 1 & 1.42 &  1  & \tiny{infeasible}  & 23 & 5.98  \\ \hhline{||-||-|-|-|-||-|-|-|-||}
15 & 1 & \tiny{infeasible} & 1 &  26.55 & 1  & \tiny{infeasible} & 31 & 41.64 \\
 \hhline{|b:=:b:=:=:=:=:b:=:=:=:=:b|}
\end{tabular}
\vspace{-1mm}
\caption{Verification of $M_n$ (\textsc{T} in seconds). 
}
\vspace{-1mm}
\label{fig:results}
\end{minipage}
\end{figure}

We consider two properties: $\Phi = A (\true \,\U (x \!\geq\! 0))$
and $\Phi' = A (\true \,\U (x \!\geq\! 1))$. The property $\Phi$ is satisfied by all
variants in $\Kk$, whereas $\Phi'$ is violated only by one configuration
$\neg A_1 \!\land\! \ldots \!\land\! \neg A_n$ (where all features are disabled).
We have verified $M_n$ against $\Phi$ and $\Phi'$ using \fsmv (e.g.\ see \fsmv models for $M_1$ and $M_2$ in Fig.~\ref{fig:fsmv}, Appendix~\ref{app:example}).
We have also checked $M_n$ using our \texttt{Verify} procedure.
For $\Phi$, \texttt{Verify} terminates in one iteration since $\joinasym(M_n)$
satisfies $\Phi$ (see $G_{\joinasym(M_1) \times \Phi}$ in Fig.~\ref{fig:GGM1P0}, Appendix~\ref{app:example}).
For $\Phi'$, \texttt{Verify} needs $n+1$ iterations. 
First, an indefinite result is reported for $\joinasym(M_n)$ (e.g. see
$G_{\joinasym(M_1) \times \Phi'}$ in Fig.~\ref{fig:GGM1P1}, Appendix~\ref{app:example}), and the
configuration space is split into $\sbr{\neg A_1}$ and $\sbr{A_1}$ subsets.
The refinement procedure proceeds in this way until we obtain definite results for all variants.
The performance results are shown in Fig.~\ref{fig:results}. Notice that, \fsmv\ reports all
results in only one iteration. 
As $n$ grows, \texttt{Verify} becomes faster than \fsmv.
For $n=11$ (for which $|\Kk|=2^{11}$),
\fsmv\, timeouts after 2 hours. On the other hand, \texttt{Verify} is feasible
even for large values of $n$.

\vspace{-1mm}
\paragraph{\elevator.}
We have experimented with the \elevator\ model with four floors, designed by Plath and Ryan \cite{DBLP:journals/scp/PlathR01}.
It contains about 300 LOC of \fsmv\ code and 9 independent optional features that modify
the basic behaviour of the elevator,  thus yielding $2^9$ = 512 variants.
To use our \texttt{Verify} procedure, we have manually translated the \fsmv\ model into an FTS
and then we have called \texttt{Verify} on it.
The basic \elevator\ system consists of a single lift that travels between four floors.
There are four platform buttons and a single lift, which declares
variables $floor, door, direction$, and
a further four cabin buttons.
The lift will always serve all requests in its current direction before it stops
and changes direction.
When serving a floor, the lift door opens and closes again.
We consider three properties ``$\Phi_1=E (\ttv \, \U (floor\!=\!1 \land idle \land door\!=\!closed))$'',
``$\Phi_2=A (\ttv \, \U (floor\!=\!1 \land idle \land door\!=\!closed))$'', and
``$\Phi_3=E (\ttv \, \U ((floor\!=\!3 \land \neg liftBut3.pressed \land direction\!=\!up) \implies door\!=\!closed))$''.
The performance results are shown in Fig.~\ref{fig:casestudy:elevator}.
The properties $\Phi_1$ and $\Phi_2$ are satisfied by all variants, but \texttt{Verify} achieves speed-ups of 28 times
for $\Phi_1$ and 2.7 times for $\Phi_2$ compared to the \fsmv\ approach.
\fsmv\ takes 1.76 sec to check $\Phi_3$,
whereas \texttt{Verify} ends in 0.67 sec thus giving 2.6 times performance speed-up.

\begin{figure*}[t]
\begin{center}
\begin{tabular}{||c||c|c||c|c||c||}  \hhline{|t:=:t:=:=:t:=:=:t:=:t|}
~\emph{prop-}~ & \multicolumn{2}{c||}{\small{\fsmv}}     & \multicolumn{2}{c||}{~\small{\texttt{Verify}}~} & \multicolumn{1}{c||}{\emph{Improvement}} \\
~\emph{-erty}~  & ~\textsc{Call}~ & ~\textsc{T}~  & \textsc{Call} & ~\textsc{T}~  & ~~\textsc{Time}~~      \\ \hhline{|:=::=:=::=:=::=:|}
$\Phi_1$ & 1 & 15.22 s & 1 & 0.55 s & \phantom{1}28 $\times$ \\ \hhline{||-||--||--||-||}
$\Phi_2$ & 1 & 1.59 s & 1 & 0.59 s & \phantom{1}2.7 $\times$ \\ \hhline{||-||--||--||-||}
$\Phi_3$ & 1 & 1.76 s & 1 & 0.67 s & \phantom{1}2.6 $\times$ \\ \hhline{|b:=:b:=:=:b:=:=:b:=:b|}
\end{tabular}
\vspace{-2mm}
\caption{Verification of \elevator\, properties (\textsc{T} in seconds). 
}\label{fig:casestudy:elevator}
\end{center}
\end{figure*}

\vspace{-1mm}

\vspace{-1.5mm}

\section{Related Work and Conclusion}\label{sec:related}


There are different formalisms for representing variability models \cite{DBLP:conf/esop/LarsenNW07,Beek16}.
Classen et al. \cite{model-checking-spls} present Featured Transition Systems (FTSs), which are today widely accepted as the model
essentially sufficient for most purposes of lifted model checking. They show how
specifically designed lifted model checking algorithms~\cite{DBLP:conf/splc/CordyCHSL13,classen-model-checking-spls-icse11} can be used for verifying FTSs
against LTL and CTL properties.
The variability abstractions that preserve LTL are introduced in \cite{spin15,sttt16,DBLP:conf/birthday/DimovskiW17}, and subsequently
automatic abstraction refinement procedures \cite{DBLP:conf/sigsoft/CordyHLSDL14,fase17} for lifted model checking of LTL are proposed,
by using Craig interpolation to define the refinement.
The variability abstractions that preserve the full CTL are introduced in \cite{fase18}, but they are constructed manually and no notion of
refinement is defined there.
In this paper, we define
an automatic abstraction refinement procedure for lifted model checking of full CTL by using games to define the refinement.
To the best of our knowledge, this is the first such procedure in lifted model checking.

One of the earliest attempts for using games for CTL model checking has been proposed by Stirling~\cite{DBLP:series/txcs/Stirling01}.
Shoham and Grumberg \cite{DBLP:journals/tocl/ShohamG07,DBLP:journals/iandc/ShohamG10,DBLP:journals/iandc/GrumbergLLS07,DBLP:conf/atva/CampetelliGLT09} have
extended this game-based approach 
for CTL over 3-valued semantics. In this work, we exploit and apply the game-based 
approach in a completely new direction, for automatic CTL verification of variability models.
Thus, we view our work as establishing a brand new connection between games and SPL communities.
We hope that further interesting interplay will be possible.

The works \cite{spin16,DBLP:journals/tcs/Dimovski18} present an approach for software lifted model checking
of \texttt{\#ifdef}-based program families using symbolic game semantics models \cite{DTCS14}.

\section{Conclusion}

To conclude, in this work we present a game-based lifted model checking for abstract variability models with
respect to the full CTL. We also suggest
an automatic refinement procedure, in case the model checking result is indefinite.


\bibliographystyle{splncs03}
\bibliography{ms}

\newpage
\appendix

\section{Standard CTL Semantics}\label{app:ctl}

We formalize the standard CTL semantics over a TS $\Ttt$.
We write $\sbr{\Ttt}^{s}_{TS}$   for the set of executions that start in the state $s$; and $\rho_i=s_i$ to denote the $i$-th state of the execution $\rho = s_0 \lambda_{1} s_{1} \ldots$.

 Satisfaction of a state formula $\Phi$ in a state $s$ of a TS $\Ttt$, denoted $[\Ttt,s \models \Phi]$, is defined as:
\begin{description}
\item[(1)] $[\Ttt,s \models a] = \ttv$ iff $a \in L(s)$, \\
$[\Ttt,s \models \neg a] = \ttv$ iff $a \not\in L(s)$
\item[(2)] $[\Ttt,s \models \Phi_1 \land \Phi_2]=\ttv$ iff $[\Ttt,s \models \Phi_1]=\ttv$ and $[\Ttt,s \models \Phi_2]=\ttv$, \\
$[\Ttt,s \models \Phi_1 \lor \Phi_2]=\ttv$ iff $[\Ttt,s \models \Phi_1]=\ttv$ or $[\Ttt,s \models \Phi_2]=\ttv$
\item[(3)] $[\Ttt,s \models \forall \phi]=\ttv$ iff  $\forall \rho \in \sbr{\Ttt}^{s}_{\textrm{TS}}. \, [\Ttt,\rho \models \phi]=\ttv$; \\
$[\Ttt,s \models \exists \phi]=\ttv$ iff  $\exists \rho \in \sbr{\Ttt}^{s}_{\textrm{TS}}. \, [\Ttt,\rho \models \phi]=\ttv$
\end{description}
Satisfaction of a path formula $\phi$ for an (infinite) execution $\rho= s_0 \lambda_{1} s_{1} \ldots$ of a TS $\Ttt$, denoted $[\Ttt,\rho \models \phi]$, is
defined as:
\begin{description}
\item[(4)] $[\Ttt, \rho \models \bigcirc \Phi]=\ttv$ iff $[\Ttt, \rho_1 \models \Phi]$, \\
$[\Ttt, \rho \models (\Phi_1 \U \Phi_2)]=\ttv$ iff $\exists i \!\geq\! 0. \, \big( [\Ttt, \rho_i \models \Phi_2]=\ttv \land (\forall 0 \!\leq\! j \!\leq\! i\!-\!1. \, [\Ttt, \rho_j \models \Phi_1]=\ttv) \big)$, \\
$[\Ttt, \rho \models (\Phi_1 \V \Phi_2)]=\ttv$ iff $\forall i \!\geq\! 0. \, \big( \forall 0 \!\leq\! j \!\leq\! i\!-\!1. \, [\Ttt, \rho_j \models \Phi_1]=\ffv \implies \, [\Ttt, \rho_i \models \Phi_2]=\ttv \big)$
\end{description}
A TS $\Ttt=(S,Act,trans,I,AP,L)$ satisfies a state formula $\Phi$, written $[\Ttt \models \Phi]=\ttv$, iff all its initial states satisfy the formula:
$\forall s_0 \in I. \, [\Ttt, s_0 \models \Phi]=\ttv$.

We now define the cases of 3-valued semantics of CTL over an MTSs
 $\Mmm$, which were not given in Section~\ref{sec:background} due to page restrictions.
 A CTL state formula $\Phi$ is satisfied in a state $s$ of an MTS $\Mmm$, denoted $[\Mmm,s \models^3 \Phi]$,
iff:
\begin{description}
\item[(2)] $[s \models^3 \Phi_1 \lor \Phi_2] = \left.\begin{cases} \ttv, & \text{if } [s \models^3 \Phi_1]=\ttv \text{ or } [s \models^3 \Phi_2]=\ttv \\ \ffv, & \text{if } [s \models^3 \Phi_1]=\ffv \text{ and } [s \models^3 \Phi_2]=\ffv \\ \bot, & \text{otherwise} \end{cases}\right.$
\end{description}
Satisfaction of a path formula $\phi$ for a may- or must-execution $\rho = s_0 \lambda_{1} s_{1} \lambda_{2} \ldots$ of an MTS $\Mmm$, denoted $[\Mmm,\rho \models^3 \phi]$, is
defined as:
\begin{description}
\item[(4)] $[\rho \models^3 \bigcirc \Phi] = \left.\begin{cases} [\rho_1 \models^3 \Phi], & \text{if } |\rho|>1\\ \bot, & \text{otherwise} \end{cases}\right.$
\item[] $[\rho \models^3 \!(\Phi_1 \V \Phi_2)] \!=\! \left.\begin{cases} \ttv, & \!\!\text{if } \begin{array}{@{} l @{}} \forall 0 \!\leq\! i \!\leq\! |\rho|. \big( \forall j \!<\! i. [\rho_j \!\models^3\! \Phi_1] \!\neq\! tt \!\!\implies\!\! [\rho_i \!\models^3\! \Phi_2]\!=\!\ttv \big) \\ \land \ \forall i \!\geq\! 0. [\rho_i \!\models^3 \! \Phi_1] \!\neq\! \ttv \!\implies\!\! |\rho|=\infty \end{array}  \\[1.1ex] \ffv, & \text{if } \exists i \!\geq\! 0. \, \big( [\rho_i \models^3 \Phi_2]=\ffv \land (\forall j \!<\! i. \, [\rho_j \models^3 \Phi_1]=\ffv) \big) \\[1.1ex] \bot, & \text{otherwise} \end{cases}\right.$
\end{description} 

\newpage
\section{Proofs}\label{app:proofs}

We use the following Lemma proved in \cite{fase18} using the definitions of
$\pi_k(\Fff)$, $\joinasym(\Fff)$, $\joinasym(\psi)$, and $\widetilde{\joinasym}(\psi)$
for an FTS $\Fff$, a configuration $k \in \Kk$, and $\psi \in \FeatExp(\Ff)$

\noindent
\textbf{Lemma 2.}
\begin{description}
\item[(i)] Let $k \in \Kk$ and $\rho \in \sbr{\pi_k(\Fff)}_{TS}$. Then, $\rho \in \sbr{\joinasym(\Fff)}^{\may}_{MTS}$.
\item[(ii)] Let $\rho \in \sbr{\joinasym(\Fff)}^{\must}_{MTS}$. Then,  $\rho \in \sbr{\pi_k(\Fff)}_{TS}$ for all $k \in \Kk$.
\end{description}

\noindent
\textbf{Theorem 1.}[Preservation results]
For every $\Phi \in CTL$, we have:
\begin{description}
\item[(1)]
$[\joinasym(\Fff) \models^3 \Phi]=\ttv \ \implies \ [\Fff \models \Phi]=\ttv$.
\item[(2)]
$[\joinasym(\Fff) \models^3 \Phi]\!=\!\ffv \, \implies \, [\Fff \models \Phi]\!=\!\ffv$ and $[\pi_k(\Fff) \models \Phi]\!=\!\ffv$ for all $k \in \Kk$.
\end{description}
\begin{proof}[Theorem\,\ref{theorem:sound}]
By induction on the structure of $\Phi$.
All cases except $\forall$ and $\exists$ quantifiers are straightforward.

\noindent
Consider the case (\textbf{1}): $[\joinasym(\Fff) \models^3 \Phi]=\ttv \ \implies \ [\Fff \models \Phi]=\ttv$.

Case $\Phi=A \phi$. To prove (\textbf{1}), we proceed by contraposition.
Assume that $[\Fff \models A \phi] \neq \ttv$. Then, there exists a configuration $k \in \Kk$
and an execution $\rho \in \sbr{\pi_k(\Fff)}_{TS}$, such that $[\pi_k(\Fff),\rho \models \phi] \neq \ttv$, i.e.\ $[\pi_k(\Fff),\rho \models \neg \phi]=\ttv$.
By Lemma~2(i), we have that $\rho \in \sbr{\joinasym(\Fff)}^{\may}_{MTS}$. Thus, $[\joinasym(\Fff),\rho \models^3 \phi] \neq \ttv$, and so
$[\joinasym(\Fff) \models^3 A \phi] \neq \ttv$ by definition.

Case $\Phi=E \phi$. To prove (\textbf{1}), we
assume $[\joinasym(\Fff) \models^3 E \phi]=\ttv$.
This means that there exists an execution $\rho \in \sbr{\joinasym(\Fff)}^{\must}_{MTS}$ such that $[\joinasym(\Fff),\rho \models^3 \phi]=\ttv$.
By Lemma~2(ii), we have that for all $k \in \Kk$, it holds
$\rho \in \sbr{\pi_{k}(\Fff)}_{TS}$. Therefore,
$[\pi_k(\Fff) \models E \phi]=\ttv$ for all $k \in \Kk$, and so $[\Fff \models E \phi]=\ttv$.

$\newline$
\noindent
Consider the case (\textbf{2}): $[\joinasym(\Fff) \models^3 \Phi]=\ffv \ \implies \ [\Fff \models \Phi]=\ffv$.

Case $\Phi=A \phi$. To prove (\textbf{2}), we
assume $[\joinasym(\Fff) \models^3 A \phi]=\ffv$.
This means that there exists an execution $\rho \in \sbr{\joinasym(\Fff)}^{\must}_{MTS}$ such that $[\joinasym(\Fff),\rho \models^3 \phi]=\ffv$.
By Lemma~2(ii), we have that for all $k \in \Kk$, it holds
$\rho \in \sbr{\pi_{k}(\Fff)}_{TS}$. Therefore,
$[\pi_k(\Fff) \models A \phi]=\ffv$ for all $k \in \Kk$, and so $[\Fff \models A \phi]=\ffv$.

Case $\Phi=E \phi$. To prove (\textbf{2}), we proceed by contraposition.
Assume that $[\Fff \models E \phi] \neq \ffv$. Then, there exists a configuration $k \in \Kk$
and an execution $\rho \in \sbr{\pi_k(\Fff)}_{TS}$, such that $[\pi_k(\Fff),\rho \models \phi] \neq \ffv$, i.e.\ $[\pi_k(\Fff),\rho \models \phi]=\ttv$.
By Lemma~2(i), we have that $\rho \in \sbr{\joinasym(\Fff)}^{\may}_{MTS}$. Thus, $[\joinasym(\Fff),\rho \models^3 \phi] \neq \ffv$, and so
$[\joinasym(\Fff) \models^3 E \phi] \neq \ffv$ by definition.
\qed
\end{proof}

\noindent
\textbf{Theorem 3.}
The procedure \texttt{Verify($\Fff,\Kk,\Phi$)} terminates and is correct.
\begin{proof}[Theorem\,\ref{theorem:correct}]
At the end of an iteration, the procedure \texttt{Verify($\Fff,\Kk,\Phi$)} either terminates with answers `\ttv' or '\ffv', or
an indefinite result is returned.
In the latter case, let $\psi$ be the feature expression guarding the transition in $\Fff$ that is found as the reason for failure.
This (failure) transition occurs as a may-transition in $\joinasym(\Fff)$.
We generate $\Fff_1=\pi_{\sbr{\psi}}(\Fff)$ and $\Kk_1 = \Kk \cap \sbr{\psi}$, as well as $\Fff_2=\pi_{\sbr{\neg \psi}}(\Fff)$ and
$\Kk_2 = \Kk \cap \sbr{\neg \psi}$.
In the next iteration, we call \texttt{Verify($\Fff_1,\Kk_1,\Phi$)} and \texttt{Verify($\Fff_2,\Kk_2,\Phi$)}.
We have that $\Kk_1 \subseteq \Kk$, $\Kk_2 \subseteq \Kk$, and $\joinasym(\Fff_1)$ contains the (failure) may-transition
as a must-transition, while $\joinasym(\Fff_2)$ does not contain the (failure) may-transition at all.
In this way, we have eliminated the reason for failure in the previous iteration, since only may-transitions can be
failure reasons according to Lemma~\ref{lemma:reason}.
Given that the number of possible updates of the configuration space
and the number of may-transitions in abstract models are finite,
the number of iterations is also finite. 

If the procedure \texttt{Verify($\pi_{\Kk'}(\Fff),\Kk',\Phi$)}  terminates with answer `\ttv' that a property is satisfied
(resp., `\ffv' that a property is violated)
for the variants $k \in \Kk'$, then the answer is correct by Theorem~\ref{theorem:sound}, case (\textbf{1}) (resp., case (\textbf{2})). \qed
\end{proof}

\newpage
\section{Figures for \vending\ and $\Phi_1$}\label{app:figures}

\begin{figure}
\centering
\begin{minipage}[b]{.5\textwidth}
\centering
\begin{tikzpicture}[->,>=stealth',shorten >=0.9pt,auto,node distance=1.4cm, semithick]
  \tikzstyle{every state}=[minimum size=.2pt,initial text={{}}]

  \node[initial above,state] (A)     {\tiny $s_0$};
  \node[state]         (B) [right of=A] {\tiny $s_1$};
  \node[state]         (C) [right of=B] {\tiny $s_2$};

  \path[font=\tiny]
        (B) edge node {$\textit{drink}$} (C)
        (C) edge [bend left] node {$\textit{take}$} (A);

  \path[draw, dashed, font=\tiny]
  (A) edge node {$\textit{pay}$} (B)
  (A) edge [bend left=75]  node[above,sloped] {{$\textit{free}$}} (C);

\end{tikzpicture}
\vspace{-2.55mm}
\caption{\small $\joinasym(\pi_{\sbr{\neg \fc}}(\vending))$}
\label{fig:abstractFTSnc}
\end{minipage}%
\begin{minipage}[b]{.5\textwidth}
\centering
\begin{tikzpicture}[->,>=stealth',shorten >=0.9pt,auto,node distance=1.4cm, semithick]
  \tikzstyle{every state}=[minimum size=.2pt,initial text={{}}]

  \node[initial above,state] (A)     {\tiny $s_0$};
  \node[state]         (B) [right of=A] {\tiny $s_1$};
  \node[state]         (C) [right of=B] {\tiny $s_2$};

  \path[font=\tiny]
        (B) edge node {$\textit{drink}$} (C)
        (C) edge [bend left] node {$\textit{take}$} (A)
        (B) edge [bend right=45] node[above,sloped] {$\textit{cancel}$} (A);

  \path[draw, dashed, font=\tiny]
  (A) edge node {$\textit{pay}$} (B)
  (A) edge [bend left=75]  node[above,sloped] {{$\textit{free}$}} (C);

\end{tikzpicture}
\vspace{-2.55mm}
\caption{\small $\joinasym(\pi_{\sbr{\fc}}(\vending))$}
\label{fig:abstractFTScc}
\end{minipage}%
\end{figure}

\begin{figure}
\centering
\begin{minipage}[b]{0.9\textwidth}
\centering
\begin{tikzpicture}[->,>=stealth',level distance=2.9em,sibling distance=5.8em,level 1/.style={sibling distance=5.8em},
    level 2/.style={sibling distance=6.2em},level 2/.style={sibling distance=6.4em}]
  \node[blue](0) {\tiny $(s_0,A(\neg \sr \U \sr))$}
    child { node[blue] {\tiny $(s_0,\sr \lor (\neg \sr \land A \bigcirc A (\neg \sr \U \sr)))$} edge from parent[standard]
      child { node[red,label=left:{\tiny $Q1$}] {\tiny $(s_0, \sr)$} edge from parent[standard] }
      child { node[blue] {\tiny $(s_0,\neg \sr \land A \bigcirc A (\neg \sr \U \sr))$} edge from parent[standard]
        child { node[blue,label=left:{\tiny $Q2$}] {\tiny $(s_0,\neg \sr)$} edge from parent[standard] }
        child { node[blue] {\tiny $(s_0,A \bigcirc A (\neg \sr \U \sr))$} edge from parent[standard]
                child { node[blue,label=above:{\textit{pay}}] {\tiny $(s_1,A (\neg \sr \U \sr))$} edge from parent[dash]
                        child { node[blue] {\tiny $(s_1,\sr \lor (\neg \sr \land A \bigcirc A (\neg \sr \U \sr)))$} edge from parent[standard]
                                   child { node[red,label=left:{\tiny $Q3$}] {\tiny $(s_1, \sr)$} edge from parent[standard] }
                                   child { node[blue] {\tiny $(s_1,\neg \sr \land A \bigcirc A (\neg \sr \U \sr))$} edge from parent[standard]
                                        child { node[blue,label=left:{\tiny $Q4$}] {\tiny $(s_1,\neg \sr)$} edge from parent[standard] }
                                        child  { node[blue] {\tiny $(s_1,A \bigcirc A (\neg \sr \U \sr))$} edge from parent[standard] }
                                   }
                        }
                        child {edge from parent[draw=none]}
                }
                child { node[blue,label=above:{\textit{free}}] {\tiny $(s_2,A (\neg \sr \U \sr))$} edge from parent[dash]
                }
            } }
                                     };


       \path[draw]
            (0-1-2-2-1-1-2-2) edge  [bend right=45] node[left] {$\textit{drink}$} (0-1-2-2-2);
\end{tikzpicture}
\end{minipage}
\caption{The colored game-graph for $\joinasym(\pi_{\sbr{\neg \fc}}(\vending))$ and $\Phi_1=A(\neg \sr \U \sr)$.}
\label{fig:refine2-1}
\end{figure}

\begin{figure}
\centering
\begin{minipage}[b]{.5\textwidth}
\centering
\begin{tikzpicture}[->,>=stealth',shorten >=0.9pt,auto,node distance=1.4cm, semithick]
  \tikzstyle{every state}=[minimum size=.2pt,initial text={}]

  \node[initial above,state] (A)     {\tiny $s_0$};
  \node[state]         (B) [right of=A] {\tiny $s_1$};
  \node[state]         (C) [right of=B] {\tiny $s_2$};

  \path[font=\tiny]
        (A) edge node {$\textit{pay}$} (B)
        (B) edge [bend right=45] node[above,sloped] {$\textit{cancel}$} (A)
        (B) edge node {$\textit{drink}$} (C)
        (C) edge [bend left] node {$\textit{take}$} (A);

\end{tikzpicture}
\vspace{-2.5mm}
\caption{\small $\joinasym(\pi_{\sbr{\fc \land \neg \ff}}(\vending))$}
\label{fig:abstractFTScnf}
\end{minipage}%
\begin{minipage}[b]{.5\textwidth}
\centering
\begin{tikzpicture}[->,>=stealth',shorten >=0.9pt,auto,node distance=1.4cm, semithick]
  \tikzstyle{every state}=[minimum size=.2pt,initial text={}]

  \node[initial above,state] (A)     {\tiny $s_0$};
  \node[state]         (B) [right of=A] {\tiny $s_1$};
  \node[state]         (C) [right of=B] {\tiny $s_2$};

  \path[font=\tiny]
        (A) edge [bend left=75]  node[above,sloped] {{$\textit{free}$}} (C)
        (B) edge node {$\textit{drink}$} (C)
        (C) edge [bend left] node {$\textit{take}$} (A)
        (B) edge [bend right=45] node[above,sloped] {$\textit{cancel}$} (A);

\end{tikzpicture}
\vspace{-2.5mm}
\caption{\small $\joinasym(\pi_{\sbr{\fc \land \ff}}(\vending))$}
\label{fig:abstractFTScf}
\end{minipage}%
\end{figure}

\begin{figure}
\centering
\begin{minipage}[b]{0.575\textwidth}
\hspace{-2cm}
\centering
\begin{tikzpicture}[->,>=stealth',level distance=2.9em,sibling distance=5.8em,level 1/.style={sibling distance=5.8em},
    level 2/.style={sibling distance=6.2em},level 2/.style={sibling distance=6.4em}]
  \node[red](0) {\tiny $(s_0,A(\neg \sr \U \sr))$}
    child { node[red] {\tiny $(s_0,\sr \lor (\neg \sr \land A \bigcirc A (\neg \sr \U \sr)))$} edge from parent[standard]
      child { node[red,label=left:{\tiny $Q1$}] {\tiny $(s_0, \sr)$} edge from parent[standard] }
      child { node[red] {\tiny $(s_0,\neg \sr \land A \bigcirc A (\neg \sr \U \sr))$} edge from parent[standard]
        child { node[blue,label=left:{\tiny $Q2$}] {\tiny $(s_0,\neg \sr)$} edge from parent[standard] }
        child { node[red] {\tiny $(s_0,A \bigcirc A (\neg \sr \U \sr))$} edge from parent[standard]
                child { node[red,label=above:{\textit{pay}}] {\tiny $(s_1,A (\neg \sr \U \sr))$} edge from parent[standard]
                        child { node[red] {\tiny $(s_1,\sr \lor (\neg \sr \land A \bigcirc A (\neg \sr \U \sr)))$} edge from parent[standard]
                                   child { node[red,label=above:{\tiny $Q3$}] {\tiny $(s_1, \sr)$} edge from parent[standard] }
                                   child { node[red] {\tiny $(s_1,\neg \sr \land A \bigcirc A (\neg \sr \U \sr))$} edge from parent[standard]
                                        child { node[blue,label=above:{\tiny $Q4$}] {\tiny $(s_1,\neg \sr)$} edge from parent[standard] }
                                        child  { node[red] {\tiny $(s_1,A \bigcirc A (\neg \sr \U \sr))$} edge from parent[standard] }
                                   }
                        }
                        child {edge from parent[draw=none]}
                }
                child { node[blue] {\tiny $(s_2,A (\neg \sr \U \sr))$} edge from parent[draw=none]
                }
            } }
                                     };


       \path[draw]
            (0-1-2-2-1-1-2-2) edge [relative, out=90, in=80, looseness=1.491] node[right] {$\textit{cancel}$} (0);
       \path[draw]
            (0-1-2-2-1-1-2-2) edge [bend right=45] node[left] {$\textit{drink}$} (0-1-2-2-2);
\end{tikzpicture}
\vspace{-8mm}
\caption{$G_{\joinasym(\pi_{\sbr{\fc \land \neg \ff}}(\vending)) \times \Phi_1}$}
\label{fig:refine3-1}
\end{minipage}%
\begin{minipage}[b]{.425\textwidth}
\centering
\hspace{1cm}
\centering
\begin{tikzpicture}[->,>=stealth',level distance=2.9em,sibling distance=5.8em,level 1/.style={sibling distance=5.8em},
    level 2/.style={sibling distance=6.2em},level 2/.style={sibling distance=6.4em}]
  \node[blue](0) {\tiny $(s_0,A(\neg \sr \U \sr))$}
    child { node[blue] {\tiny $(s_0,\sr \lor (\neg \sr \land A \bigcirc A (\neg \sr \U \sr)))$} edge from parent[standard]
      child { node[red,label=left:{\tiny $Q1$}] {\tiny $(s_0, \sr)$} edge from parent[standard] }
      child { node[blue] {\tiny $(s_0,\neg \sr \land A \bigcirc A (\neg \sr \U \sr))$} edge from parent[standard]
        child { node[blue,label=left:{\tiny $Q2$}] {\tiny $(s_0,\neg \sr)$} edge from parent[standard] }
        child { node[blue] {\tiny $(s_0,A \bigcirc A (\neg \sr \U \sr))$} edge from parent[standard]
                child { node[blue,label=above right:{\textit{free}}] {\tiny $(s_2,A (\neg \sr \U \sr))$} edge from parent[standard]
                }
            } }
                                     };


\end{tikzpicture}
\vspace{-0mm}
\caption{$G_{\joinasym(\pi_{\sbr{\fc \land \ff}}(\vending)) \times \Phi_1}$}
\label{fig:refine3-2}
\end{minipage}%
\end{figure} 

\newpage
\section{Figures for \vending\ and $\Phi_2$}\label{app:figures2}

\begin{figure}
\centering
\hspace{-2cm}
\vspace{-2mm}
\begin{minipage}[b]{0.9\textwidth}
\centering
\begin{tikzpicture}[->,>=stealth',level distance=2.95em,sibling distance=5.8em,level 1/.style={sibling distance=5.8em},
    level 2/.style={sibling distance=6.2em},level 2/.style={sibling distance=6.4em}]
  \node[white](0) {\tiny $(s_0,E(\neg \textit{r} \U \textit{r}))$}
    child { node[white] {\tiny $(s_0,\textit{r} \lor (\neg \textit{r} \land E \bigcirc E (\neg \textit{r} \U \textit{r})))$} edge from parent[standard]
      child { node[red,label=left:{\tiny $Q1$}] {\tiny $(s_0, \sr)$} edge from parent[standard] }
      child { node[white] {\tiny $(s_0,\neg \sr \land E \bigcirc E (\neg \sr \U \sr))$} edge from parent[standard]
        child { node[blue,label=left:{\tiny $Q2$}] {\tiny $(s_0,\neg \sr)$} edge from parent[standard] }
        child { node[white,label=below:{\tiny failure node}] {\tiny $(s_0,E \bigcirc E (\neg \sr \U \sr))$} edge from parent[standard]
                child { node[blue,label=above:{\textit{pay}}] {\tiny $(s_1,E (\neg \sr \U \sr))$} edge from parent[dash]
                        child { node[blue] {\tiny $(s_1,\sr \lor (\neg \sr \land E \bigcirc E (\neg \sr \U \sr)))$} edge from parent[standard]
                                   child { node[red,label=left:{\tiny $Q3$}] {\tiny $(s_1, \sr)$} edge from parent[standard] }
                                   child { node[blue] {\tiny $(s_1,\neg \sr \land E \bigcirc E (\neg \sr \U \sr))$} edge from parent[standard]
                                        child { node[blue,label=above:{\tiny $Q4$}] {\tiny $(s_1,\neg \sr)$} edge from parent[standard] }
                                        child  { node[blue] {\tiny $(s_1,E \bigcirc E (\neg \sr \U \sr))$} edge from parent[standard] }
                                   }
                        }
                        child {edge from parent[draw=none]}
                }
                child { node[blue,label=above:{\textit{free}}] {\tiny $(s_2,E (\neg \sr \U \sr))$} edge from parent[dash]
                        child {edge from parent[draw=none]}
                        child { node[blue] {\tiny $(s_2,\sr \lor (\neg \sr \land E \bigcirc E (\neg \sr \U \sr)))$} edge from parent[standard]
                                   child { node[blue,label=above:{\tiny $Q5$}] {\tiny $(s_2, \sr)$} edge from parent[standard] }
                                   child { node[red] {\tiny $(s_2,\neg \sr \land E \bigcirc E (\neg \sr \U \sr))$} edge from parent[standard]
                                        child { node[red,label=above:{\tiny $Q6$}] {\tiny $(s_2,\neg \sr)$} edge from parent[standard] }
                                        child { node[white] {\tiny $(s_2,E \bigcirc E (\neg \sr \U \sr))$} edge from parent[standard] }
                                   }
                        }
                }
            } }
                                     };

       \path[draw, dashed]
            (0-1-2-2-1-1-2-2) edge [relative, out=90, in=80, looseness=1.491] node[right] {$\textit{cancel}$} (0);
       \path[draw]
            (0-1-2-2-1-1-2-2) edge node[left] {$\textit{drink}$} (0-1-2-2-2)
            (0-1-2-2-2-2-2-2) [bend right=70] edge node[left] {$\textit{take}$} (0);
\end{tikzpicture}
\end{minipage}
\vspace{-9.75mm}
\caption{The colored game-graph for $\joinasym(\pi_{\sbr{\neg \fc}}(\vending))$ and $\Phi_2=E(\neg \sr \U \sr)$.
The failure node is $(s_0,E \bigcirc E (\neg \sr \U \sr))$, and the reason is the may-edge
$(s_0,E \bigcirc E (\neg \sr \U \sr)) \to (s_2, E (\neg \sr \U \sr))$.
The corresponding transition in \vending\ is $s_0 \TR{\textit{free}/\!f} s_2$.
In the next iteration, consider $\pi_{\sbr{f}}(\vending)$ and $\pi_{\sbr{\neg f}}(\vending)$.}
\label{fig2:refine2-1}
\end{figure}

\begin{figure}
\centering
\begin{minipage}[b]{.5\textwidth}
\centering
\begin{tikzpicture}[->,>=stealth',shorten >=0.9pt,auto,node distance=1.4cm, semithick]
  \tikzstyle{every state}=[minimum size=.2pt,initial text={}]

  \node[initial above,state] (A)     {\tiny $s_0$};
  \node[state]         (B) [right of=A] {\tiny $s_1$};
  \node[state]         (C) [right of=B] {\tiny $s_2$};

  \path[font=\tiny]
        (A) edge [bend left=65]  node[above,sloped] {{$\textit{free}$}} (C)
        (B) edge node {$\textit{drink}$} (C)
        (C) edge [bend left] node {$\textit{take}$} (A);
  \path[draw, dashed, font=\tiny]
        (B) edge [bend right=45] node[above,sloped] {$\textit{cancel}$} (A);

\end{tikzpicture}
\vspace{-2.5mm}
\caption{\small $\joinasym(\pi_{\sbr{\ff}}(\vending))$}
\label{fig2:abstractFTScf}
\end{minipage}%
\begin{minipage}[b]{.5\textwidth}
\centering
\begin{tikzpicture}[->,>=stealth',shorten >=0.9pt,auto,node distance=1.4cm, semithick]
  \tikzstyle{every state}=[minimum size=.2pt,initial text={}]

  \node[initial above,state] (A)     {\tiny $s_0$};
  \node[state]         (B) [right of=A] {\tiny $s_1$};
  \node[state]         (C) [right of=B] {\tiny $s_2$};

  \path[font=\tiny]
        (A) edge node {$\textit{pay}$} (B)
        (B) edge node {$\textit{drink}$} (C)
        (C) edge [bend left] node {$\textit{take}$} (A);

  \path[draw, dashed, font=\tiny]
        (B) edge [bend right=45] node[above,sloped] {$\textit{cancel}$} (A);

\end{tikzpicture}
\vspace{-2.5mm}
\caption{\small $\joinasym(\pi_{\sbr{\neg \ff}}(\vending))$}
\label{fig2:abstractFTScnf}
\end{minipage}%
\end{figure}

\begin{figure}
\centering
\hspace{-2cm}
\centering
\begin{tikzpicture}[->,>=stealth',level distance=2.95em,sibling distance=5.8em,level 1/.style={sibling distance=5.8em},
    level 2/.style={sibling distance=6.2em},level 2/.style={sibling distance=6.4em}]
  \node[blue](0) {\tiny $(s_0,E(\neg \textit{r} \U \textit{r}))$}
    child { node[blue] {\tiny $(s_0,\textit{r} \lor (\neg \textit{r} \land E \bigcirc E (\neg \textit{r} \U \textit{r})))$} edge from parent[standard]
      child { node[red,label=left:{\tiny $Q1$}] {\tiny $(s_0, \sr)$} edge from parent[standard] }
      child { node[blue] {\tiny $(s_0,\neg \sr \land E \bigcirc E (\neg \sr \U \sr))$} edge from parent[standard]
        child { node[blue,label=left:{\tiny $Q2$}] {\tiny $(s_0,\neg \sr)$} edge from parent[standard] }
        child { node[blue] {\tiny $(s_0,E \bigcirc E (\neg \sr \U \sr))$} edge from parent[standard]
                child { node[blue,label=above right:{\textit{free}}] {\tiny $(s_2,E (\neg \sr \U \sr))$} edge from parent[standard]
                }
            } }
                                     };

\end{tikzpicture}
\vspace{-1mm}
\caption{The colored game-graph for $\joinasym(\pi_{\sbr{\ff}}(\vending))$ and $\Phi_2=E(\neg \sr \U \sr)$.
All variants from $\sbr{\ff}$ satisfy $\Phi_2$.}
\label{fig2:refine3-1}
\end{figure}

\vspace{-5mm}

\begin{figure}
\centering
\hspace{-1cm}
\centering
\begin{tikzpicture}[->,>=stealth',level distance=2.95em,sibling distance=5.8em,level 1/.style={sibling distance=5.8em},
    level 2/.style={sibling distance=6.2em},level 2/.style={sibling distance=6.4em}]
  \node[blue](0) {\tiny $(s_0,E(\neg \textit{r} \U \textit{r}))$}
    child { node[blue] {\tiny $(s_0,\textit{r} \lor (\neg \textit{r} \land E \bigcirc E (\neg \textit{r} \U \textit{r})))$} edge from parent[standard]
      child { node[red,label=left:{\tiny $Q1$}] {\tiny $(s_0, \sr)$} edge from parent[standard] }
      child { node[blue] {\tiny $(s_0,\neg \sr \land E \bigcirc E (\neg \sr \U \sr))$} edge from parent[standard]
        child { node[blue,label=left:{\tiny $Q2$}] {\tiny $(s_0,\neg \sr)$} edge from parent[standard] }
        child { node[blue] {\tiny $(s_0,E \bigcirc E (\neg \sr \U \sr))$} edge from parent[standard]
                child { node[blue,label=above right:{\textit{pay}}] {\tiny $(s_1,E (\neg \sr \U \sr))$} edge from parent[standard]
                        child {edge from parent[draw=none]}
                }
            } }
                                     };

\end{tikzpicture}
\vspace{-0mm}
\caption{The colored game-graph for $\joinasym(\pi_{\sbr{\neg \ff}}(\vending))$ and $\Phi_2=E(\neg \sr \U \sr)$.
All variants from $\sbr{\neg \ff}$ satisfy $\Phi_2$.}
\label{fig2:refine3-2}
\end{figure} 

\newpage
\section{Example}\label{app:example}

\begin{figure}
\begin{subfigure}[b]{.48\linewidth}
\centering
$ \begin{array}{l}
\textcolor{gray}{1}~~ \texttt{MODULE \emph{features} }\\
\textcolor{gray}{2}~~ \texttt{VAR} \  fA1 : boolean;\\
\textcolor{gray}{3}~~ \texttt{ASSIGN}\\
\textcolor{gray}{4}~~ \quad \texttt{init}(fA1) := \{\textit{TRUE,FALSE}\}; \\
\textcolor{gray}{5}~~ \quad \texttt{next}(fA1) := fA1; \\
\textcolor{gray}{6}~~ \texttt{MODULE \emph{main} } \\
\textcolor{gray}{7}~~  \texttt{VAR}\ f:features; \, x:0..1; \, nA1 : 0..1;\\
\textcolor{gray}{8}~~  \texttt{ASSIGN}\\
\textcolor{gray}{9}~~  \ \texttt{init}(x) := 0; \\
\textcolor{gray}{10}~~  \texttt{init}(nA1) := 0; \\
\textcolor{gray}{11}~~ \texttt{next}(x) := \texttt{case} \\
\textcolor{gray}{12}~~ \qquad \qquad \qquad f.fA1 \, \& \, nA1\!=\!0 : x\!+\!1; \\
\textcolor{gray}{13}~~ \qquad \qquad \qquad \textit{TRUE : } x; \\
\textcolor{gray}{14}~~ \qquad \qquad \quad \texttt{easc}; \\
\textcolor{gray}{15}~~ \texttt{next}(nA1) := \texttt{case} \\
\textcolor{gray}{16}~~ \quad  f.fA1 \, \& \, nA1\!=\!0 \, \& \, \texttt{next}(x)\!=\!x\!+\!1: 1; \\
\textcolor{gray}{17}~~ \quad  \textit{TRUE : } nA1; \\
\textcolor{gray}{18}~~ \qquad \qquad \qquad \, \texttt{easc}; \\
\textcolor{gray}{19}~~ \texttt{SPEC} \ AF (x \geq 1);
\end{array} $
\vspace{-1mm}
\caption{\fsmv model for $M_1$.}
\label{fig:fsmv1}
\end{subfigure}
\hfill
\begin{subfigure}[b]{.48\linewidth}
\centering
$ \begin{array}{l}
\textcolor{gray}{1}~~ \texttt{MODULE \emph{features} }\\
\textcolor{gray}{2}~~ \texttt{VAR} \  fA1 : boolean; \, fA2 : boolean;\\
\textcolor{gray}{3}~~ \texttt{ASSIGN}\\
\textcolor{gray}{4}~~ \quad \texttt{init}(fA1) := \{\textit{TRUE,FALSE}\}; \\
\textcolor{gray}{5}~~ \quad \texttt{init}(fA2) := \{\textit{TRUE,FALSE}\}; \\
\textcolor{gray}{6}~~ \quad \texttt{next}(fA1) := fA1; \\
\textcolor{gray}{7}~~ \quad \texttt{next}(fA2) := fA2; \\
\textcolor{gray}{8}~~ \texttt{MODULE \emph{main} } \\
\textcolor{gray}{9}~~  \texttt{VAR} \ \ f:features; \, x:0..3; \\
\textcolor{gray}{10}~~  \qquad nA1 : 0..1; \, nA2 : 0..1; \\
\textcolor{gray}{11}~~  \texttt{ASSIGN}\\
\textcolor{gray}{12}~~  \ \texttt{init}(x) := 0;  \\
\textcolor{gray}{13}~~  \ \texttt{init}(nA1) := 0; \\
\textcolor{gray}{14}~~  \ \texttt{init}(nA2) := 0; \\
\textcolor{gray}{15}~~ \ \texttt{next}(x) := \texttt{case} \\
\textcolor{gray}{16}~~ \qquad \qquad \qquad f.fA1 \, \& \, nA1\!=\!0 : x\!+\!1; \\
\textcolor{gray}{17}~~ \qquad \qquad \qquad f.fA2 \, \& \, nA2\!=\!0 : x\!+\!2; \\
\textcolor{gray}{18}~~ \qquad \qquad \qquad \textit{TRUE : } x; \\
\textcolor{gray}{19}~~ \qquad \qquad \quad \ \texttt{easc}; \\
\textcolor{gray}{20}~~ \texttt{next}(nA1) := \texttt{case} \\
\textcolor{gray}{21}~~ \quad  f.fA1 \, \& \, nA1\!=\!0 \, \& \, \texttt{next}(x)\!=\!x\!+\!1: 1; \\
\textcolor{gray}{22}~~ \quad  \textit{TRUE : } nA1; \\
\textcolor{gray}{23}~~ \qquad \qquad \qquad \ \texttt{easc}; \\
\textcolor{gray}{24}~~ \texttt{next}(nA2) := \texttt{case} \\
\textcolor{gray}{25}~~ \quad  f.fA2 \, \& \, nA2\!=\!0 \, \& \, \texttt{next}(x)\!=\!x\!+\!2: 1; \\
\textcolor{gray}{26}~~ \quad  \textit{TRUE : } nA2; \\
\textcolor{gray}{27}~~ \qquad \qquad \qquad \ \texttt{easc}; \\
\textcolor{gray}{28}~~ \texttt{SPEC} \ AF (x \geq 1);
\end{array} $
\vspace{-1mm}
\caption{\fsmv model for $M_2$.}
\label{fig:fsmv2}
\end{subfigure}
\vspace{-2mm}
\caption{\fsmv\ models.} \label{fig:fsmv}
\end{figure}

\begin{figure}[t]
\centering
\centering
\hspace{-2cm}
\begin{tikzpicture}[->,>=stealth',level distance=3em,sibling distance=6.5em,level 1/.style={sibling distance=6.5em},
    level 2/.style={sibling distance=7.0em},level 2/.style={sibling distance=7.4em}]
  \node[blue,label=right:{\tiny $Q12$}](0) {\tiny $(s_0,A(\ttv \,\U (x \!\geq\! 0)))$}
    child { node[blue,label=right:{\tiny $Q11$}] {\tiny $(s_0,(x \!\geq\! 0) \lor (\ttv \land A \bigcirc A (\ttv \,\U (x \!\geq\! 0))))$} edge from parent[standard]
      child { node[blue,label=left:{\tiny $Q1$}] {\tiny $(s_0, x \!\geq\! 0)$} edge from parent[standard] }
      child { node[blue,label=right:{\tiny $Q10$}] {\tiny $(s_0,\ttv \land A \bigcirc A (\ttv \,\U (x \!\geq\! 0)))$} edge from parent[standard]
        child { node[blue,label=left:{\tiny $Q2$}] {\tiny $(s_0,\ttv)$} edge from parent[standard] }
        child { node[blue,label=right:{\tiny $Q9$}] {\tiny $(s_0,A \bigcirc A (\ttv \,\U (x \!\geq\! 0)))$} edge from parent[standard]
                child { node[blue,label=above :{$\tau$},label=left:{\tiny $Q7$}] {\tiny $(s_1,A(\ttv \,\U (x \!\geq\! 0)))$} edge from parent[dash]
                        child { node[blue] {\tiny $(s_1,(x \!\geq\! 0) \lor (\ttv \land A \bigcirc A (\ttv \,\U (x \!\geq\! 0))))$} edge from parent[standard]
                                   child { node[blue,label=left:{\tiny $Q3$}] {\tiny $(s_1, x \!\geq\! 0)$} edge from parent[standard] }
                                   child { node[blue] {\tiny $(s_1,\ttv \land A \bigcirc A (\ttv \,\U (x \!\geq\! 0)))$} edge from parent[standard]
                                        child { node[blue,label=above:{\tiny $Q4$}] {\tiny $(s_1,\ttv)$} edge from parent[standard] }
                                        child  { node[blue] {\tiny $(s_1,A \bigcirc A (\ttv \U (x \!\geq\! 0)))$} edge from parent[standard] }
                                   }
                        }
                        child {edge from parent[draw=none]}
                }
                child { node[blue,label=above :{$\tau$},label=right:{\tiny $Q8$}] {\tiny $(s_2,A(\ttv \,\U (x \!\geq\! 0)))$} edge from parent[dash]
                        child {edge from parent[draw=none]}
                        child { node[blue] {\tiny $(s_2,(x \geq 0) \lor (\ttv \land A \bigcirc A (\ttv \,\U (x \!\geq\! 0))))$} edge from parent[standard]
                                   child { node[blue,label=above:{\tiny $Q5$}] {\tiny $(s_2, x \!\geq\! 0)$} edge from parent[standard] }
                                   child { node[blue] {\tiny $(s_2,\ttv \land A \bigcirc A (\ttv \,\U (x \!\geq\! 0)))$} edge from parent[standard]
                                        child { node[blue,label=above:{\tiny $Q6$}] {\tiny $(s_2,\ttv)$} edge from parent[standard] }
                                        child { node[blue] {\tiny $(s_2,A \bigcirc A (\ttv \,\U (x \!\geq\! 0)))$} edge from parent[standard] }
                                   }
                        }
                }
            } }
                                     };

       \path[draw]
            (0-1-2-2-1-1-2-2) edge [relative, out=90, in=50, looseness=4.0] node[right] {$\tau$} (0-1-2-2-1);
       \path[draw]
            (0-1-2-2-2-2-2-2) edge [bend right=95] node[right] {$\tau$} (0-1-2-2-2);
\end{tikzpicture}
\vspace{-8mm}
\caption{The colored game-graph for $\joinasym(M_1)$ (Fig~\ref{fig:FTS-joinM1}) and $\Phi=A(\ttv \,\U (x \!\geq\! 0))$.
Note that $Q7$ (resp., $Q8$) is the may-MSCC derived using $s_1$ and $A(\ttv \,\U (x \!\geq\! 0))$ (resp, $s_2$ and $A(\ttv \,\U (x \!\geq\! 0))$).}
\label{fig:GGM1P0}
\end{figure}
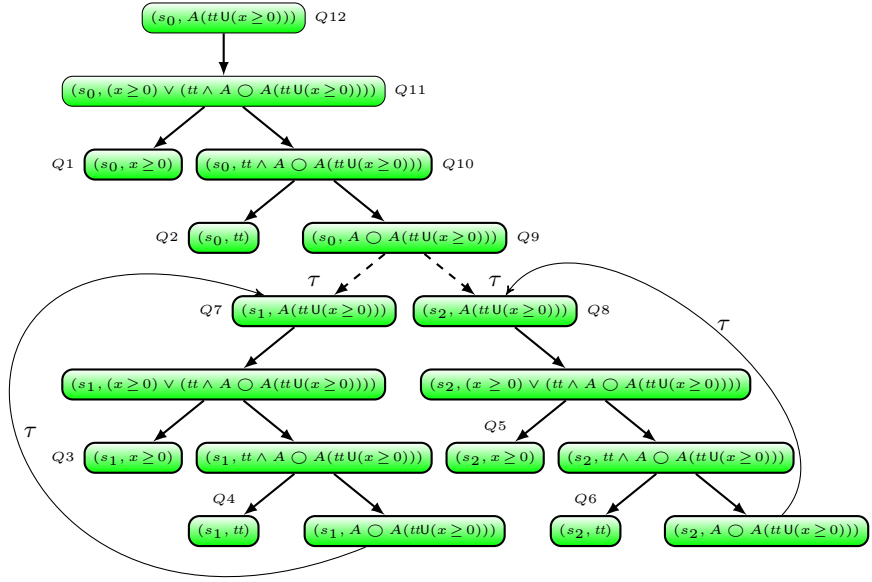

\begin{figure}
\centering
\begin{minipage}[b]{.45\textwidth}
\centering
\begin{tikzpicture}[->,>=stealth',shorten >=0.9pt,auto,node distance=1.1cm, semithick]
  \tikzstyle{every state}=[minimum size=.2pt,initial text={{}}]

  \node[initial left,state] (A)     [label=right:{\tiny $s_0$}]  {\tiny $0$};
  \node[state]         (B) [above right of=A, label=below:{\tiny $s_1$}] {\tiny $1$};
  \node[state]         (C) [below right of=A, label=below:{\tiny $s_2$}] {\tiny $0$};

  \path[draw, font=\tiny] (A) edge node[above, sloped] {$\tau /\! A_1$} (B)
                    (A) edge node[below, sloped] {$\tau /\! \neg A_1$} (C);

  \path[font=\tiny] (C) edge [in=60, out=30, looseness=9] node[above,pos=.9] {$\tau$} (C)
                    (B) edge [in=60, out=30, looseness=9] node[above,pos=.9] {$\tau$} (B);

\end{tikzpicture}
\vspace{-2.5mm}
\caption{\small The model $M_1$.}
\label{fig:FTS-M1}
\end{minipage}%
\begin{minipage}[b]{.45\textwidth}
\centering
\begin{tikzpicture}[->,>=stealth',shorten >=0.9pt,auto,node distance=1.1cm, semithick]
  \tikzstyle{every state}=[minimum size=.2pt,initial text={{}}]

  \node[initial left,state] (A)     [label=above:{\tiny $s_0$}]  {\tiny $0$};
  \node[state]         (B) [above right of=A, label=above:{\tiny $s_1$}] {\tiny $1$};
  \node[state]         (C) [below right of=A, label=below:{\tiny $s_2$}] {\tiny $0$};

  \path[draw, dashed, font=\tiny] (A) edge node[above, sloped] {} (B)
                    (A) edge node[below, sloped] {} (C);

  \path[font=\tiny] (C) edge [in=60, out=30, looseness=9] node[above,pos=.9] {} (C)
                    (B) edge [in=60, out=30, looseness=9] node[above,pos=.9] {} (B);

\end{tikzpicture}
\vspace{-2.5mm}
\caption{\small The model $\joinasym(M_1)$.}
\label{fig:FTS-joinM1}
\end{minipage}%
\end{figure}

\begin{figure}
\centering
\centering
\hspace{-2cm}
\begin{tikzpicture}[->,>=stealth',level distance=3em,sibling distance=6.5em,level 1/.style={sibling distance=6.5em},
    level 2/.style={sibling distance=7.0em},level 2/.style={sibling distance=7.4em}]
  \node[white,label=right:{\tiny $Q12$}](0) {\tiny $(s_0,A(\ttv \,\U (x \!\geq\! 1)))$}
    child { node[white,label=right:{\tiny $Q11$}] {\tiny $(s_0,(x \!\geq\! 1) \lor (\ttv \land A \bigcirc A (\ttv \,\U (x \!\geq\! 1))))$} edge from parent[standard]
      child { node[red,label=left:{\tiny $Q1$}] {\tiny $(s_0, x \!\geq\! 1)$} edge from parent[standard] }
      child { node[white,label=right:{\tiny $Q10$}] {\tiny $(s_0,\ttv \land A \bigcirc A (\ttv \,\U (x \!\geq\! 1)))$} edge from parent[standard]
        child { node[blue,label=left:{\tiny $Q2$}] {\tiny $(s_0,\ttv)$} edge from parent[standard] }
        child { node[white,label=right:{\tiny $Q9$},label=below:{\tiny failure node}] {\tiny $(s_0,A \bigcirc A (\ttv \,\U (x \!\geq\! 1)))$} edge from parent[standard]
                child { node[blue,label=above :{$\tau$},label=left:{\tiny $Q7$}] {\tiny $(s_1,A(\ttv \,\U (x \!\geq\! 1)))$} edge from parent[dash]
                        child { node[blue] {\tiny $(s_1,(x \!\geq\! 1) \lor (\ttv \land A \bigcirc A (\ttv \,\U (x \!\geq\! 1))))$} edge from parent[standard]
                                   child { node[blue,label=left:{\tiny $Q3$}] {\tiny $(s_1, x \!\geq\! 1)$} edge from parent[standard] }
                                   child { node[blue] {\tiny $(s_1,\ttv \land A \bigcirc A (\ttv \,\U (x \!\geq\! 1)))$} edge from parent[standard]
                                        child { node[blue,label=above:{\tiny $Q4$}] {\tiny $(s_1,\ttv)$} edge from parent[standard] }
                                        child  { node[blue] {\tiny $(s_1,A \bigcirc A (\ttv \,\U (x \!\geq\! 1)))$} edge from parent[standard] }
                                   }
                        }
                        child {edge from parent[draw=none]}
                }
                child { node[red,label=above :{$\tau$},label=right:{\tiny $Q8$}] {\tiny $(s_2,A(\ttv \,\U (x \!\geq\! 1)))$} edge from parent[dash]
                        child {edge from parent[draw=none]}
                        child { node[red] {\tiny $(s_2,(x \geq 1) \lor (\ttv \land A \bigcirc A (\ttv \,\U (x \!\geq\! 1))))$} edge from parent[standard]
                                   child { node[red,label=above:{\tiny $Q5$}] {\tiny $(s_2, x \!\geq\! 1)$} edge from parent[standard] }
                                   child { node[red] {\tiny $(s_2,\ttv \land A \bigcirc A (\ttv \,\U (x \!\geq\! 1)))$} edge from parent[standard]
                                        child { node[blue,label=above:{\tiny $Q6$}] {\tiny $(s_2,\ttv)$} edge from parent[standard] }
                                        child { node[red] {\tiny $(s_2,A \bigcirc A (\ttv \,\U (x \!\geq\! 1)))$} edge from parent[standard] }
                                   }
                        }
                }
            } }
                                     };

       \path[draw]
            (0-1-2-2-1-1-2-2) edge [relative, out=90, in=50, looseness=4.0] node[right] {$\tau$} (0-1-2-2-1);
       \path[draw]
            (0-1-2-2-2-2-2-2) edge [bend right=95] node[right] {$\tau$} (0-1-2-2-2);
\end{tikzpicture}
\vspace{-8mm}
\caption{The colored game-graph for $\joinasym(M_1)$ (Fig~\ref{fig:FTS-joinM1}) and $\Phi'=A(\ttv \,\U (x \!\geq\! 1))$.
Note that $Q7$ (resp., $Q8$) is the may-MSCC derived using $s_1$ and $A(\ttv \,\U (x \!\geq\! 1))$ (resp, $s_2$ and $A(\ttv \,\U (x \!\geq\! 1))$).
The failure node is $(s_0,A \bigcirc A (\ttv \U (x \!\geq\! 1)))$, and the reason is the may-edge
$(s_0,A \bigcirc A (\ttv \U (x \!\geq\! 1))) \to (s_2, A (\ttv \U (x \!\geq\! 1)))$.
The corresponding concrete transition in $M_1$ is $s_0 \TR{\tau/\!\neg A_1} s_2$.
So, in the next third iteration we consider FTSs $\pi_{\sbr{A_1}}(M_1)$ and $\pi_{\sbr{\neg A_1}}(M_1)$.}
\label{fig:GGM1P1}
\end{figure}
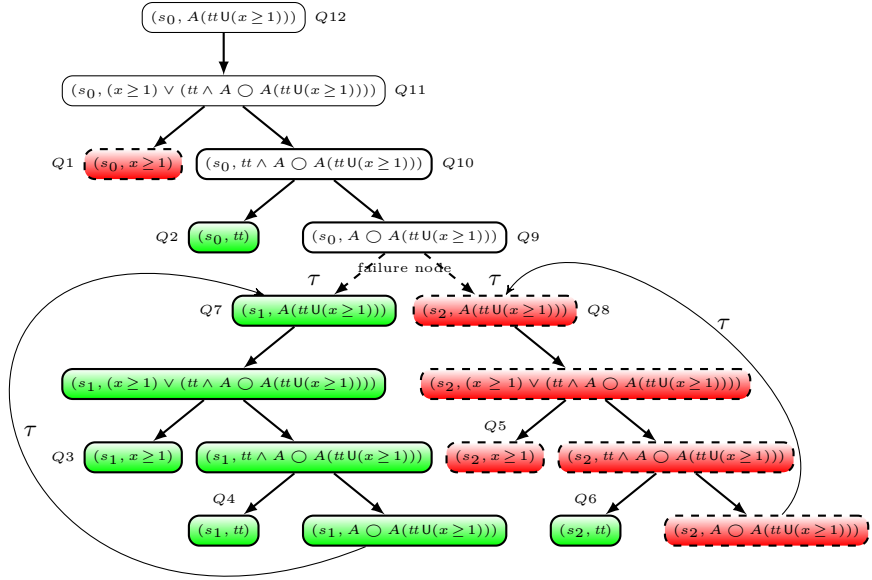 

\end{document}